\documentclass[sigconf]{acmart}

\usepackage{booktabs} % For formal tables
\usepackage[T1]{fontenc}
\usepackage{graphicx}
\usepackage[normalem]{ulem}
\usepackage{slantsc}
\usepackage{amsmath}
\usepackage{amssymb}
\usepackage{fancybox}
\definecolor{jean}{HTML}{188E20}
\usepackage{mdframed}
\usepackage{listings}
\usepackage{colortbl}
\usepackage{array}
\usepackage{natbib}
\usepackage{multirow}
\usepackage{multicol}
\usepackage{listings}
\usepackage{subfloat}
\usepackage[captionskip=1pt,farskip=1pt,position=below]{subfig}
\usepackage{boxedminipage}
\usepackage[ruled,vlined,lined,commentsnumbered]{algorithm2e}
\usepackage{balance}
\usepackage{csvsimple}
\usepackage{longtable}
\usepackage{booktabs}
\usepackage[normalem]{ulem}
\usepackage[skip=1pt,captionskip=1pt,labelfont=bf]{caption}

\newcommand{\tthat}{\mbox{\tt \char'136}}

\usepackage{multirow}
\usepackage{amsmath}
\usepackage{amssymb}
\usepackage{ifthen}
\usepackage{pgfplots}
\usepackage{color}
\usepackage{colortbl}
\usepackage{multirow}
\usepackage{pgfplotstable}

\usepackage[most]{tcolorbox}

\tcbset{beforeafter skip=2pt,top=2pt,left=2pt,right=2pt,bottom=2pt}

\usetikzlibrary{external}
\usetikzlibrary{patterns}
\usetikzlibrary{pgfplots.groupplots}

\usepackage{alltt}

\usepackage{threeparttable}

\usepackage{marvosym}

\usepackage{enumitem}
\setlist{noitemsep} %to leave space around whole list

\definecolor{codegray}{gray}{0.9}

\definecolor{yellow}{RGB}{255,255,153}
\definecolor{grey}{RGB}{224,224,224}
\definecolor{burntorange}{RGB}{0.8, 0.33, 0.0}

\newboolean{showcomments}
\setboolean{showcomments}{true}
\ifthenelse{\boolean{showcomments}}
 { \newcommand{\mynote}[2]{
      \fbox{\bfseries\sffamily\scriptsize#1}
        {\small$\blacktriangleright$\textsf{\emph{#2}}$\blacktriangleleft$}}}
        { \newcommand{\mynote}[2]{}}

\newcommand{\patternfix}{{\sc Par}\xspace}
\newcommand{\genprog}{GenProg\xspace}

\newcommand{\manual}{H\xspace}

\copyrightyear{2017} 
\acmYear{2017} 
\setcopyright{acmcopyright}
\acmConference{ISSTA '17}{July 10-14, 2017}{Santa Barbara, CA, USA}\acmPrice{15.00}\acmDOI{10.1145/3092703.3092713}
\acmISBN{978-1-4503-5076-1/17/07}

\acmConference[ISSTA'17]{International Symposium on Software Testing and Analysis}{July 2017}{Santa Barbara, California USA} 
\acmYear{2017}
\copyrightyear{2017}

\begin{document}

\title{Impact of Tool Support in Patch Construction}
%\subtitle{A study of Linux  Patches}
%\titlenote{Produces the permission block, and
%  copyright information}
%\subtitle{Extended Abstract}
%\subtitlenote{The full version of the author's guide is available as
%  \texttt{acmart.pdf} document}

\author{Anil Koyuncu}
\affiliation{%
  \institution{SnT, University of Luxembourg - Luxembourg}
}
  
\author{Tegawend\'e F. Bissyand\'e}
\affiliation{%
  \institution{SnT, University of Luxembourg - Luxembourg}
}
  
\author{Dongsun Kim}
\affiliation{%
  \institution{SnT, University of Luxembourg - Luxembourg}
}
  
\author{Jacques Klein}
\affiliation{%
  \institution{SnT, University of Luxembourg - Luxembourg}
}
  
\author{Martin Monperrus}
\affiliation{%
  \institution{Inria, University of Lille - France}
}  
\author{Yves Le Traon}
\affiliation{%
  \institution{SnT, University of Luxembourg - Luxembourg}
}
\renewcommand{\shortauthors}{Anil Koyuncu, Tegawend\'e F. Bissyand\'e, Dongsun Kim,\\ Jacques Klein, Martin Monperrus, Yves Le Traon}

%\vspace{-8mm} 
\begin{abstract}
In this work, we investigate the practice of patch construction in the Linux kernel development, 
focusing on the differences between three patching processes: (1) patches crafted entirely manually to fix bugs, 
(2) those that are derived from warnings of bug detection tools, and (3) those that
are automatically generated based on fix patterns. With this study, we 
provide to the research community concrete insights on the practice
of patching as well as how the development community is currently embracing research and commercial patching tools to improve productivity in repair. 
%In particular, we investigate the extent of the 
%acceptance of bug finding and patch application tools in a production environment, and study the opportunities of automation
%that the automated repair community can explore. 
%Practically, our investigations provide insights
%to researchers on the  acceptance of tool-supported patches. 
%\dongsun{I have replaced a sentence here to stress out our finding. Please check.}
The result of our study shows that tool-supported patches are increasingly adopted by the developer community while manually-written patches are accepted more quickly. 
Patch application tools enable developers to remain committed to contributing patches to the code base.
%In particular, developers with extensive knowledge tend to write patches with an aid of bug detection tools. Fix patterns largely help less-experienced developers to keep contributing patches to the code base.
Our findings also include that, in actual development processes, patches generally implement several change operations spread over the code, even for patches fixing warnings by bug detection tools. Finally, this study has shown that there is an opportunity to directly leverage the output of bug detection tools to readily generate patches that are appropriate for fixing the problem, and that are consistent with manually-written patches.% in the code base.
%On the other hand, the findings in this study provide an opportunity for the community
%to reflect on which types of bugs should be a priority for automated repair.
%\dongsun{later, we need to put more specific findings with concrete results from our investigation.}
\end{abstract}

\keywords{Repair, Debugging, Patch, Linux, Empirical, Tools, Automation}
%\vspace{-0.5cm}

\begin{CCSXML}
<ccs2012>
<concept>
<concept_id>10011007.10011074.10011111.10011695</concept_id>
<concept_desc>Software and its engineering~Software version control</concept_desc>
<concept_significance>300</concept_significance>
</concept>
<concept>
<concept_id>10011007.10011006.10011073</concept_id>
<concept_desc>Software and its engineering~Software maintenance tools</concept_desc>
<concept_significance>500</concept_significance>
</concept>
<concept>
<concept_id>10011007.10011006.10011071</concept_id>
<concept_desc>Software and its engineering~Software configuration management and version control systems</concept_desc>
<concept_significance>300</concept_significance>
</concept>
</ccs2012>
\end{CCSXML}

\ccsdesc[500]{Software and its engineering~Software maintenance tools}
\ccsdesc[300]{Software and its engineering~Software configuration management and version control systems}
\ccsdesc[300]{Software and its engineering~Software version control}

\maketitle

%\vspace{-6mm} 
\section{Introduction}
\label{sec.introduction}

Patch construction is a key task in software development. In particular, it is central
to the repair process when developers must engineer change operations for fixing the buggy code. In recent years, a number of tools have been integrated into software development ecosystems, contributing to reducing the burden of patch construction. 
The process of a patch construction indeed includes various steps that can more or less be automated: bug detection tools for example can help human developers characterize and often localize the piece of code to fix, while patch application tools can systematize the formation of concrete patches that can be applied within an identified context of the code. 

Tool support however can impact patch construction in a way that may influence acceptance 
or that focuses the patches to specific bug kinds. The growing field of automated repair\cite{le_goues_genprog:_2012,kim_automatic_2013,mechtaev_angelix:_2016,nguyen_semfix:_2013}, for example, is currently challenged by the nature of the patches that are produced and their eventual acceptance by development teams. Indeed, constructed patches must be applied to a code base and later maintained by human developers.

%Automated program repair is a growing field of software engineering that promises
%to substantially alleviate the costs of debugging, which is one of the most tedious and 
%time-consuming tasks in software development~\cite{Bissyande2012,joyce89}.
%There has been a large number of techniques proposed by the community, 
%which can broadly be classified into 
%two categories: generate-and-validate~\cite{le_goues_genprog:_2012,kim_automatic_2013} and satisfiability-based~\cite{mechtaev_angelix:_2016,nguyen_semfix:_2013} repair approaches.
%% we can remove followings if it looks quite verbose.
%Techniques in the fist category mutate the code until a repair action is found that makes the program pass all test cases in the test suite. 
%On the other hand, those in the second category
%rely, for example, on symbolic execution of the program to identify the range of acceptable 
%values and make appropriate changes to fix the program. Eventually, the produced patch must be applied
%to a code base and manually maintained by human developers. 

This situation raises the question
of the acceptance of patches within a development team, with regards to the process that was relied upon to construct them. The goal of our study is therefore to identify different types of patches written by different construction processes by exploring patches in a real-world project,
to reflect on how program repair is conducted in current development settings. In particular, 
we investigate how advances in static bug detection and patch application have 
already been exploited to reduce human efforts in repair. 

We formulate research questions for 
comparing different types of patches, produced with varying degrees of automation, to offer to
the community some insights on i) whether tool-supported patches can be readily adopted, ii) whether tool-supported patches target specific kinds of bugs, and iii) where further opportunities lie for improving automated repair techniques in production environments.

In this work, we consider the Linux operating system development since it has established 
an important code base in the history of software engineering. Linux is furthermore a
reliable artifact~\cite{Israeli:JSS10} for research as patches are validated by a strongly hierarchical community before they can reach the mainline code base. Developers involved in Linux development, especially maintainers who are in charge of acknowledging patches, have relatively extensive experience
in programming. Linux's development history constitutes a valuable information for repair studies as a number of tools have been introduced in this community to automate and systematize various tasks such as code style checking, bug detections, and systematic patching.
Our analysis unfolds as an empirical comparative study 
of three patch construction processes:
\begin{itemize}[leftmargin=*]
	\item {\bf Process H:} In the first process, developers must rely on a bug report written by a user to understand the problem, locate the faulty part of source code, and manually craft a fix. We refer to it as {\em Process H}, since all steps in the process appear to involve {\bf H}uman intervention.
	\item {\bf Process DLH:} In the second process, static analysis tools first scan the source code and report on lines which are likely faulty. Fixing the reported lines of code can be straightforward since the tools may be very descriptive on the nature of the problem. Nevertheless, dealing with static debugging tools can be tedious for developers with little experience as these tools often yield too many false positives. We refer to this process as {\em Process DLH}, since {\bf D}etection and {\bf L}ocalization are automated but {\bf H}uman intervention is required to form the patch. 
	\item {\bf Process HMG:} Finally, in the third process, developers may rely on a systematic patching tool to search for and fix a specific bug pattern. We refer to this process as {\em Process HMG}, since {\bf H}uman input is needed to express the bug/fix patterns which are {\bf M}atched by a tool to a code base to {\bf G}enerate a concrete patch.
\end{itemize}

We ensure that the collected dataset does not include patch instances that can be attributed to more than one of the processes described above. Our analyses have eventually yielded a few implications for future research:

%\begin{enumerate}[leftmargin=*]
%	\item
	
	\noindent
	 {\bf \em Acceptance of patches:} development communities, such as the Linux kernel team, are becoming aware of the potential of tool support in patch construction i) to gain time by prioritizing engineering tasks and ii) to attract contributions from novice developers seeking to join a project.

\noindent
 {\bf \em  Kinds of bugs:} Tool-supported patches do not target the same kinds of bugs as manual patches. However, we note that patches fixing warnings outputted by bug detection tools are already complex, requiring several change operations over several lines, hunks and even files of code. 
 
 \noindent
{\bf \em Opportunities for automated repair:} We have performed preliminary analyses which show that bug detection tools can be leveraged as a stepping stone for automated repair in conjunction with patch generation tools, to produce patches that are consistent with human patches (for maintenance), correct (derived from past experience of fixing a specific bug type) and thus likely to be rapidly accepted by development teams. 
%\end{enumerate}

%\vspace{-3mm} 
\section{Background}
\label{sec.background}
\label{sec:linux}
%\subsection{Basics of Linux}
Linux is an open-source operating system that is widely used in
environments ranging from embedded systems to servers.  The heart of the
Linux operating system is the Linux kernel, which comprises all the code
that runs with kernel privileges, including device drivers and file
systems.  It was  first
introduced in 1994, and has grown to 14.3 million lines of C code with the
release of Linux 4.8 in Oct. 2016.\footnote{Computed with David A. Wheeler's `SLOCCount'.}  All data used in this paper are related
to changes propagated to the mainline code base until Oct. 2, 2016\footnote{Kernel's Git HEAD commit id is {\tt c8d2bc9bc39ebea8437fd974fdbc21847bb897a3}.}.

% the following is optional. Let's see paper length.
A recent study has shown that, for a collection of typical types
of faults in C code, the number of faults is staying stable, even though
the size of the kernel is increasing, implying that the overall quality of
the code is improving~\cite{Palix:2011:FLT:1950365.1950401}.  Nevertheless, ensuring the
correctness and maintainability of the code remains an important issue for Linux developers, as
reflected by discussions on the kernel mailing list~\cite{lkml}.

%\subsubsection*{\bf Development Model of Linux}

The Linux kernel is developed according to a hierarchical open source model
referred to as Benevolent dictator for life
(BDFL)~\cite{bdfl},
in which anyone can contribute, but ultimately all contributions are
integrated by a single person, Linus Torvalds.  A Linux kernel maintainer
receives patches related to a particular file or subsystem from developers
or more specialized maintainers. After evaluating and locally committing them,
he/she propagates them upwards in the maintainer hierarchy eventually towards Linus Torvalds.

%Linux's guidelines for submitting patches recommend the use of specific tags
%in the changelogs to designate developer and maintainer involvement.
%The ``{\tt Signed-off-by:}'' tag indicates that the signer was involved in the
%development of the patch, or that he/she was in the patch's delivery path. In contrast,
%if a person was not directly involved in the preparation or handling of a
%patch but wishes to signify and record their approval of it then they can
%ask to have an ``{\tt Acked-by:}'' line added to the patch's changelog.

Finally, Linux developers are urged to ``solve a single problem per patch''\footnote{see {\tt Documentation/SubmittingPatches} in linux tree.},
and maintainers are known to enforce this rule as revealed by discussions on contributors' patches
 in the Linux Kernel Mailing List (LKML)~\cite{lkml} archive.

%\subsubsection*{\bf Patching and Repair in Linux}
Recently, the development and
maintenance of the Linux kernel have become a massive effort, involving a huge
number of people. 1,731 distinct commit authors have contributed to the
development of Linux 4.8\footnote{Obtained using \texttt{git log
		v4.7..v4.8 | grep {\tthat}Author | sort -u | wc -l}, without
	controlling for variations in names or email addresses.}.  The patches
written by these commit authors are then validated by the 1,142
{\em maintainers} of Linux 4.8\footnote{Obtained using \texttt{grep
		{\tthat}M: MAINTAINERS | sort -u | wc -l} without controlling for
	variations in names or email addresses.}, who are responsible for the
various subsystems.

Since the release of Linux 2.6.12 in June 2005, the
Linux kernel has used the source code management system {\tt
	git}~\cite{git}. The current Linux kernel git
tree~\cite{linuxKernelGit}
only goes back to Linux 2.6.12, and thus we use this version as the
starting point of our study.  Between Linux 2.6.12 and Linux 4.8 there
were 616,291 commits, by 20,591 different developers\footnote{Again, we
	have not controlled for variations in names or email addresses.}.  These
commits are retrievable from the git repository as {\em patches}.
Basically, a patch is an extract of code, in which lines
beginning with \verb+-+ are to be removed lines beginning with \verb-+-
are to be added.

% the following is optional. Most of ISSTA people of course know patches and commits in git and s/w development.
%A patch
%is a line-by-line textual representation of the additions and removals of
%code made by the commit.  Figure~\ref{fig:patch} shows an example of patches. Basically, a patch is an extract of code, in which lines
%beginning with \verb+-+ are to be removed an lines beginning with \verb-+-
%are to be added.

%\dongsun{key is commit message. we can ommit commit diff in Figure 1.}
%\begin{center}
%\begin{figure}[!h]
%{\parbox{\linewidth}{%
%	\lstinputlisting[linewidth={\linewidth},basicstyle=\footnotesize\ttfamily,frame=tb]{fig/patch-tehuti.list}
%}}%
%\caption{Example of patch on the file tehuti.c.}
%\label{fig:patch}
%\end{figure}
%\end{center}

The Linux kernel community actively uses the Bugzilla~\cite{bugzilla}
issue tracking system to report and manage bugs. As of November 2016, over 28 thousands bug reports
were filed in the kernel tracking system, with about 6,000 marked as highly severe or even blocking.

The Linux community has also built, or integrated, a number of tools for improving the quality of its source code in
a systematic way. For example, The mainline code base includes the coding style checker {\em checkpatch},
which was released in July 2007, in Linux 2.6.22. The use of checkpatch is supported by the Linux kernel
guidelines for submitting patches\footnote{{\tt Documentation/SubmittingPatches} in the Linux tree.}, and checkpatch has been regularly maintained and extended since its inception. Sparse~\cite{sparse} is another example of the tools built by Linus Torvalds and colleagues to enforce typechecking.

Commercial tools, such as Coverity~\cite{coverity}, also often help to fix Linux code.
More recently, researchers at Inria/LiP6 have developed the Coccinelle project~\cite{coccinelle} for
Linux code matching and transformation. Initially, the project was designed to help developers perform collateral evolutions~\cite{padioleau08}. It is now intensively used by Linux developers to apply fix patterns to the whole code base.

%\vspace{-6mm}
\section{Methodology}
\label{sec.dataset}
Our objective is to empirically check the impact of tool support in the patch construction process in Linux. To achieve this goal,
we must collect a large, consistent and clean set of patches constructed in different processes.
Specifically, we require:

\begin{itemize}[leftmargin=*]
	\item[(1)] patches that have been a-priori manually prepared by developers
based on the knowledge of a potential bug, somewhere in the code. For this
type of patches, we assume that a user may have reported an issue while running the code.
In the Linux ecosystem, such reporters are often kernel developers.
	\item[(2)] patches that have been constructed by using the output of bug finding tools, which are integrated into the
development chain. We consider this type of patches to be tool-supported, as debugging tools
often provide reliable information on what the bug is (hence, how to fix it) and where it
is located.
	\item[(3)] patches that have been constructed, by a tool, based fully on change rules. Such fixes, validated by maintainers, are actually
based on templates of fix patterns which are used to i) match (i.e., locate) incorrect code in the project and
ii) generate a corresponding concrete fix.
\end{itemize}
%\vspace{-6mm}
\subsection{Dataset Collection}
To collect patches constructed via Process H, hereafter referred to as {\bf \em H patches}, we consider patches whose commits are explicitly
linked to a bug report from the kernel bugzilla tracking system and any other Linux distributions bug tracking systems.
We consider that such patches have been engineered manually after a careful consideration of
the report filed by a user, and often after a replication step where developers dynamically test the software.

Until Linux 4.8, we have found 5,758 patches fixing defects described in bug reports. Unfortunately,
for some of the patches, the link to its bug report provided in the commit log was not accessible (e.g., because of restriction in access rights  of some Redhat bug reports or because the web page was no longer live). Consequently, we
were able  to collect 4,417 bug patches corresponding to a
bug report~(i.e.,~$\sim77\%$~of H~patches).
Table~\ref{tab:manual} provides statistics on the bugs associated with those patches.

\begin{table}[!htb]
\centering
		\scriptsize

\caption{Statistics on H patches in Linux Kernel.}% (Bug reports and their corresponding patches for each severity level).}
\label{tab:manual}
%\subfloat[]
%{\parbox{0.4\linewidth}{%
		\begin{tabular}{lrr}
			\hline
			Sevirity & \# reports & \# patches \\
			\toprule
			Severe & 965 & 1,052 \\
			Medium & 2,961 & 3,163 \\
			Minor & 138 & 136 \\
			Enhancement & 47 & 66 \\ \hline
			Total & 4,111 & 4,417 \\
			\bottomrule
		\end{tabular}
%}}%
%\hspace{0.1in}%
% The following is just for recording.
%-\multicolumn{2}{c|}{Bug reports} & Reporters & Patches & Commit authors  & Files\\
%-\toprule
%- Severe:  &965& \multirow{4}{*}{2434} & {\#: 1052} & \multirow{2}{*}{\#: 1088}  & \multirow{2}{*}{\#: 9650} \\
%- Medium:  &2961 & &{\#: 3163}  &  & \\
%- Minor:  &138 & & {\#: 136} & \multirow{2}{*}{\%: 6.95}  & \multirow{2}{*}{\%: 17,38}  \\
%- Enhancement: &47 & & {\#: 66} &  & \\
%Total: &4111 & &{\#: 4417/5758}  &15654  &55499 \\
\end{table}

First, we note that the severity of most bugs (2,961, i.e., 72.0\%) is medium,
and H patches have fixed substantially more severe bugs (965, i.e., 23.5\%)
than minor bugs (138, i.e., 3.3\%). Only 47 (1.1\%) bug reports represent mere enhancements. Second, exploring the data shows that
there is not always a 1 to 1 relationship between bug reports and patches: a bug
report may be addressed by several patches, while a single patch may relate to
several bug reports. Nevertheless, we note that 4,270 out of 5,265 (i.e., 89\%) patches
address a single bug report. Third, a large number of unique developers (1,088 out of 18,733= 6.95\%) have provided H patches to fix user bug reports. Finally, H patches have touched about
17\% (= 9,650/57,195) of files in the code base.
Overall, these statistics suggest that the dataset
of H patches is diverse as they are indeed written by a variety of  developers
to fix a variably severe set of bugs spread across different files of the program.

We identify patches constructed via Process DLH, hereafter referred to as {\bf \em DLH patches}, by matching in commit
logs messages on the form ``{\tt found by <tool>}''\footnote{We also use ``{\tt generated by <tool>}''  since the commit authors also often refer to warnings as ``generated by'' a given tool.} where {\tt <tool>}
refers to a tool used by kernel developers to find bugs. In this work, we consider the following notable tools, for static analysis:
%\tb{Check if LDV is static and strace is dynamic}

\begin{itemize}[leftmargin=*]
	\item {\em checkpatch}: a coding style checker for ensuring some basic level of patch quality.
	\item {\em sparse}: an in-house tool for static code analysis that helps kernel developers to detect coding errors based on developer annotations.
	\item {\em Linux driver verification (LDV) project} : a set of programs, such as the Berkeley Lazy Abstraction Software verification Tool (BLAST) that solves the reachability problem, dedicated to improving the quality of kernel driver modules.
	\item {\em Smatch}: a static analysis tool.
	\item {\em Coverity}: a commercial static analysis tool.
	\item {\em Cppcheck}: an extensible static analysis tool that feeds on checking rules to detect bugs.
\end{itemize}
and for dynamic analysis:
\begin{itemize}[leftmargin=*]
	\item {\em Strace}: a tracer for system calls and signals, to monitor interactions between processes and the Linux kernel.
	\item {\em Syzkaller}: a supervised, coverage-guided Linux syscall fuzzer for testing untrusted user input.
	\item {\em Kasan}: the Linux Kernel Address SANitizer is a dynamic memory error detector for finding use-after-free and out-of-bounds bugs.
\end{itemize}

After collecting patches referring to those tools, we further check that commit logs include terms  ``bug'' or ``fix'', to focus on bug fix patches.
Table~\ref{tab:semi-manual} provides details on the distribution of patches produced based on the
output of those tools.

%\begin{table*}[!htb]
%\centering
%\caption{Statistics on semi-manual repair data.\dongsun{should be revised to two-column and vertical table.}}
%\resizebox{\linewidth}{!}{
%\label{tab:semi-manual}
%\begin{tabular}{l c c c c c c c c c }
%\toprule
%{\bf Tool} & checkpatch & sparse & LDV  & smatch & coverity & cppcheck & strace & syzkaller & kasan \\
%\midrule
% {\bf \# patches} & 292 & 68 & 220 & 39 & 84 & 14 & 4 & 7 & 1 \\
%%{\bf \# patches} &  & & & & \\
%\bottomrule
%\end{tabular}
%}
%\end{table*}

\begin{table}[!htb]
	\centering
	\scriptsize
	\caption{Statistics on DLH patches in Linux Kernel.}% (the number of patches for each static/dynamic analysis tools used during the development).}
	\label{tab:semi-manual}
	\begin{tabular}{lr|lr}
		\toprule
		{\bf Tool} &  {\bf \# patches} & {\bf Tool} &  {\bf \# patches} \\ \midrule
		checkpatch &               292 &   sparse   &  68               \\
		LDV        &               220 &   smatch   &  39               \\
		coverity   &                84 &   cppcheck &  14               \\
		strace     &                 4 &  syzkaller &  7                \\
		kasan      &                 1 &         ~  &  ~                \\
		\bottomrule
	\end{tabular}
\end{table}

{\em Checkpatch} and the {\em Linux driver verification project} tools are the most mentioned in commit logs. The {\em Coverity} commercial tool and the {\em sparse}
internal tool also helped to find and fix dozens of bugs in the kernel. Finally, we note
that static tools are more frequently referred to than dynamic tools.

{\bf \em HMG patches} in Linux are mainly carried out by Coccinelle, which was originally
designed to document and automate collateral evolutions in the kernel source
code~\cite{padioleau08}.
%It is now used in various code bases as a tool performing
%control-flow-based program searches and transformations in C code~\cite{Brunel:2009:FFP:1480881.1480897}.
Coccinelle is built on an approach where the user guides the inference process using patterns
of code that reflect the user's understanding of the conventions and design of the target
software system~\cite{LawallDSN09WYSIWIB}.

Static analysis by Coccinelle is specified by
developers who use control-flow sensitive concrete syntax matching rules~\cite{Brunel:2009:FFP:1480881.1480897}.
Coccinelle provides a language,
SmPL\footnote{Semantic Patch Language.}, for specifying search and transformations referred to
as {\em semantic patches}. It also includes a transformation engine for performing the specified
semantic patches. To avoid confusion with semantic patches in the context of automated repair literature, we will refer to Coccinelle-generated patches as {\em SmPL patches}.

\begin{figure}[!htb]
\subfloat[Example of SmPL templates.]
{\parbox{0.4\linewidth}{%
\begin{lstlisting}[linewidth={\linewidth}, frame=tb,basicstyle=\scriptsize\ttfamily]
1 @@
2 expression E;
3 constant c;
4 type T;
5 @@
6 -kzalloc(c * sizeof(T), E)
7 +kcalloc(c, sizeof(T), E)
\end{lstlisting}
}}%
\hspace{0.1in}%
\subfloat[C code matching the template on the left. (iso-kzalloc.c).]
{\parbox{0.55\linewidth}{%
\begin{lstlisting}[linewidth={\linewidth}, frame=tb,basicstyle=\scriptsize\ttfamily]
1 void main(int i)
2 {
3
4     kzalloc(2 * sizeof(int), GFP_KERNEL);
5     kzalloc(sizeof(int) * 2, GFP_KERNEL);
6
7 }
\end{lstlisting}
}}%
\caption{Illustration of SmPL matching and patching.}
\label{fig:smpl}
\end{figure}

Figure~\ref{fig:smpl} illustrates a SmPL patch example. This SmPL patch is aimed at
changing all function calls of {\em kzalloc} to {\em kcalloc} with a reorganization of call arguments. For more details on how SmPL patches are specified, we refer the reader to the project documentation\footnote{\url{http://coccinelle.lip6.fr/documentation.php}}.
%In SmPL, each rule begins with the declaration of a collection of metavariables and then
%follows with either a C-code-like pattern specifying a match in the case of a SmPL rule
%or an ordinary Ocaml/Python code to perform arbitrary computations. In case of Figure~\ref{fig:smpl}a, the semantic
%patch contains a single unnamed\footnote{A named rule is marked with @given\_name@.} rule with
%three metavariables (lines~2-4): \texttt{E} (expression) which represents an arbitrary expression,
%\texttt{c} (constant) which represents a constant value and \texttt{T} (type) which represents any data type.
%
%Metavariables are bounded by matching the code pattern against the C source code (e.g., Fig.~\ref{fig:smpl}b).
%In a SmPL patch, context and  - lines are patterns to be matched against a codebase. \texttt{+} lines are
%to be generated and inserted in replacement of matched code. For example, the pattern fragment
%on line~6 in Figure~\ref{fig:smpl}a will be matched to both lines~4~and~5 of the C code in Figure~\ref{fig:smpl}b, taking into
%account isomorphisms in operations: type \texttt{T} is then bounded to {\tt int} and \texttt{E} to the macro {\tt GFP\_KERNEL} and the constant value being 2 in both matched cases.
Figure~\ref{fig:diff} represents the concrete Unix diff generated by Coccinelle engine and which is included
in the patch to forward to mainline maintainers.

\begin{figure}[!htb]
	{\parbox{\linewidth}{
\begin{lstlisting}[linewidth={\linewidth},frame=tb,basicstyle=\scriptsize\ttfamily]
diff =
--- iso-kzalloc.c
+++ /tmp/cocci-output-52882-062587-iso-kzalloc.c
@@ -1,7 +1,7 @@
 void main(int i)
 {
-   kzalloc(2 * sizeof(int), GFP_KERNEL);
-   kzalloc(sizeof(int) * 2, GFP_KERNEL);
+   kcalloc(2, sizeof(int), GFP_KERNEL);
+   kcalloc(2, sizeof(int), GFP_KERNEL);
 }
\end{lstlisting}
	}}%
	\caption{Patch derived from the SmPL template in Figure~\ref{fig:smpl}a.}
	\label{fig:diff}
\end{figure}

%Figure~\ref{fig:coccicheck} shows an example commit message of a bug fixing patch produced using the Coccinelle pattern matching engine.
In some cases, the fix is not directly implemented
in the SmPL patch (which is then referred to as SmPL {\em match}). Nevertheless, since each bug
pattern must be clearly defined with SmPL, the associated fix is straightforward to engineer.
Overall, we have collected 4,050 HMG patches mentioning ``coccinelle'' or ``semantic patch'' and applied to C code\footnote{We have controlled with a random subset of 100 commits that this grep-based approaches yielded indeed only relevant patches constructed by Coccinelle.}.
%Nevertheless, since each bug
%pattern must be clearly defined with SmPL, the associated fix is straightforward to engineer.
%Thus, in this study we consider all Coccinelle-related patches as HMG patches.

%\begin{figure}[!htb]
%{\parbox{\linewidth}{
%		\lstinputlisting[linewidth={\linewidth}, frame=tb]{fig/concrete-patch-example.list}
%}}%
%\caption{Commit log of a SmPL-derived concrete patch applied to the Linux kernel.}
%\label{fig:coccicheck}
%%\vspace{-1cm}
%\end{figure}

%Figure~\ref{fig:manual} shows an example of bug commit

%\subsection{Heuristics for repair patch identification}

%\vspace{-0.2cm}
\subsection{Research Questions}
\label{subsec.rqs}

We now enumerate and motivate our research questions in the context of the three processes of patch construction:

\begin{itemize}
%	\item[\bf RQ0] {\em Which types of patches, among the three considered, are more prevalent for fixing bugs?} \\This research question
%	investigates how the different types of patches are used by developers over time and across project modules. Temporal distribution of patches may shed some light on the adoption and acceptance of a repair practice by kernel maintainers. Spatial distributions on the other hand may highlight the acceptance of patches based on the type (i.e., to some extent the critical nature) of the code to repair.
%
	\item[\bf RQ1] {\em How does the developer community react to the introduction of bug detection and patch application tools?} \\With this research question, we check that the temporal distributions of patches in each patch construction process are in line with the upstream discussions for accepting patches. Such discussions may shed light on the proportions of tool-supported patches that are pushed by developers but that never get into the code base.

	\item[\bf RQ2] {\em Who is using bug detection and patch application tools?} \\In this research question, we investigate the profile of patch authors in the different patch construction processes.

	\item[\bf RQ3] {\em What is the impact of patch construction process in the stability of patches?} \\We investigate the stability, i.e., whether or not the patch is reverted after being propagated in the mainline tree, of accepted patches to highlight the
	reliability of each patch application tool within the community.

	\item[\bf RQ4] {\em Do the patch construction processes target the same kind of bugs?} \\We approximate the categorization of bugs
	with two metrics related to (1) the locality of the fixes as well as (2) the nature and number of change operators of the patch.
%	We investigate whether some
%	scenarios of repair are often limited to one line or hunk of code while another may span across several lines, methods or files. Findings on this research question may help researchers focus on or disregard repair templates for local changes.
%
%	\item[\bf RQ5] {\em What {change} operators are used for fixing bugs?} \\In this research question, we study the repair operations applied in real bug fixes and compare the mutation operators used in the different repair scenarios. The objective of this study is to assess the complexity of manual repairs in comparison with semi-manual and automated patching, and provide the automated repair community with arguments for enriching the repair space of their fix templates.
	%\item[] \tb{Dongsun thinks RQ6 and RQ7 can be left out. Let's see the findings and decide}
%	\item[\bf RQ6] {\em To what extent API knowledge is a requirement for generating patches?}\\ We investigate for which scenarios
%	of patching it appears that API knowledge is important.
%
%
%	\item[\bf RQ7] {\em What is the redundancy of repair actions across patches?} \\We check the
%	repetitiveness of code changes implemented in the different scenarios of repair. %The objective is to investigate the proportion of manual patches that bring fixes which could are also carried out by automated patching tools.
%	The objective is to investigate the similarity of semi-manual fixes to conclude on the automatability of such fixes.
%	\item[RQ5] Think about efficiency.
\end{itemize}

%\section{Empirical Study of Linux Repair Patches}
%\vspace{-0.6cm}
\section{Empirical Study Findings}
\label{sec.study}

\subsection{Descriptive Statistics on the Data}
\label{sec:stat}

We first provide statistics on how the different patch construction processes are used 
by developers over time and across project modules. Temporal distribution of patches may shed some light on the adoption of a patch construction process by kernel maintainers. Spatial distributions on the other hand may highlight the acceptance of a process based on the type (i.e., to some extent the critical nature) of the code to fix.

\begin{figure}[!htb]
	\includegraphics[width=1\linewidth]{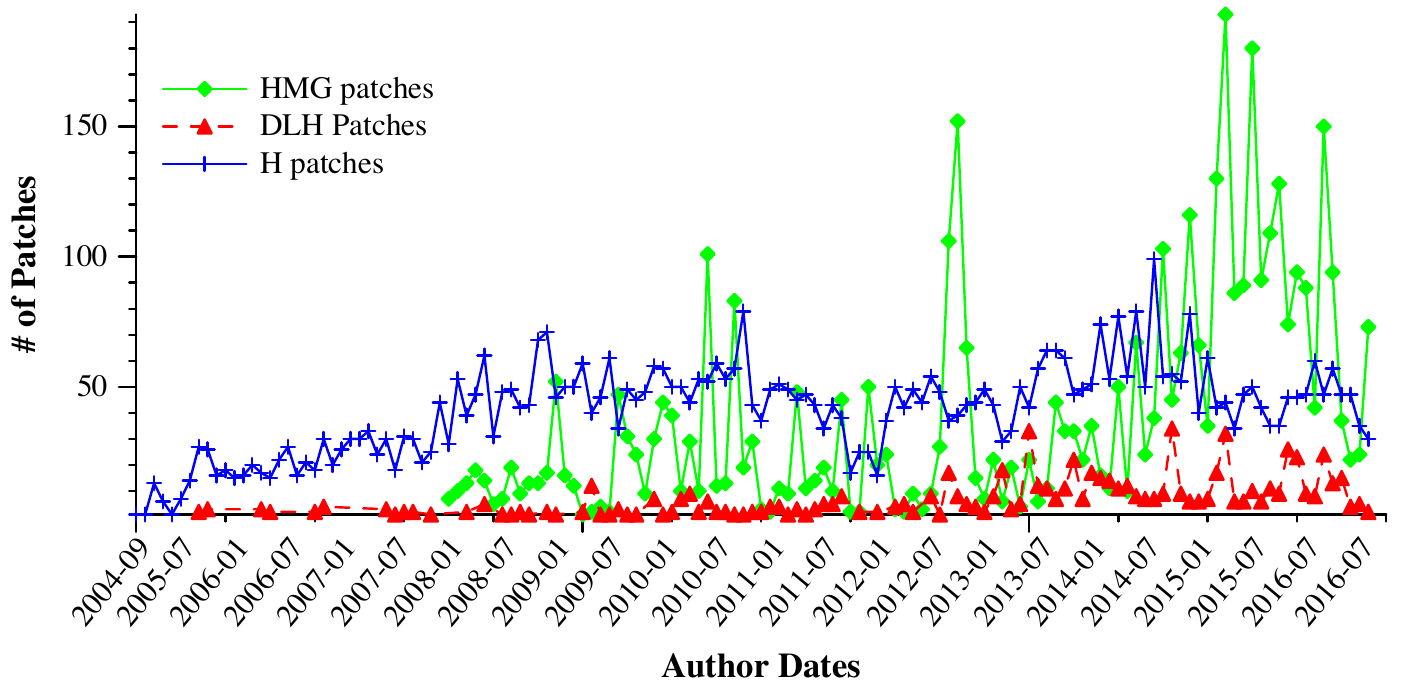}
	\caption{Temporal distributions of patches.}
	\label{fig:temporal}
\end{figure}

%\subsubsection*
{\bf Temporal distribution of patches.}
We compute the temporal distribution of patches since Linux 2.6.12 (June 2005) until 
Linux 4.8 (October 2016) and outline them in Figure~\ref{fig:temporal}. Note that 
although Linux 2.6.12 was released in June 2005, a few commit patches in the 
code base pre-date this release date.

Overall, H patches are consistently applied over time with  approximately 50  
fixes per month. DLH patches have been very slow to take up. Indeed,
the number of patches built based on bug finding tools has been narrow for 
several years, with a slight increase in recent years, partly due to the 
improvements made for reducing false positives. Finally, HMG 
patches have rapidly increased and now account for a significant portion of 
patches propagated to the mainline code base.

\begin{figure}[!htb]
	\includegraphics[width=1\linewidth]{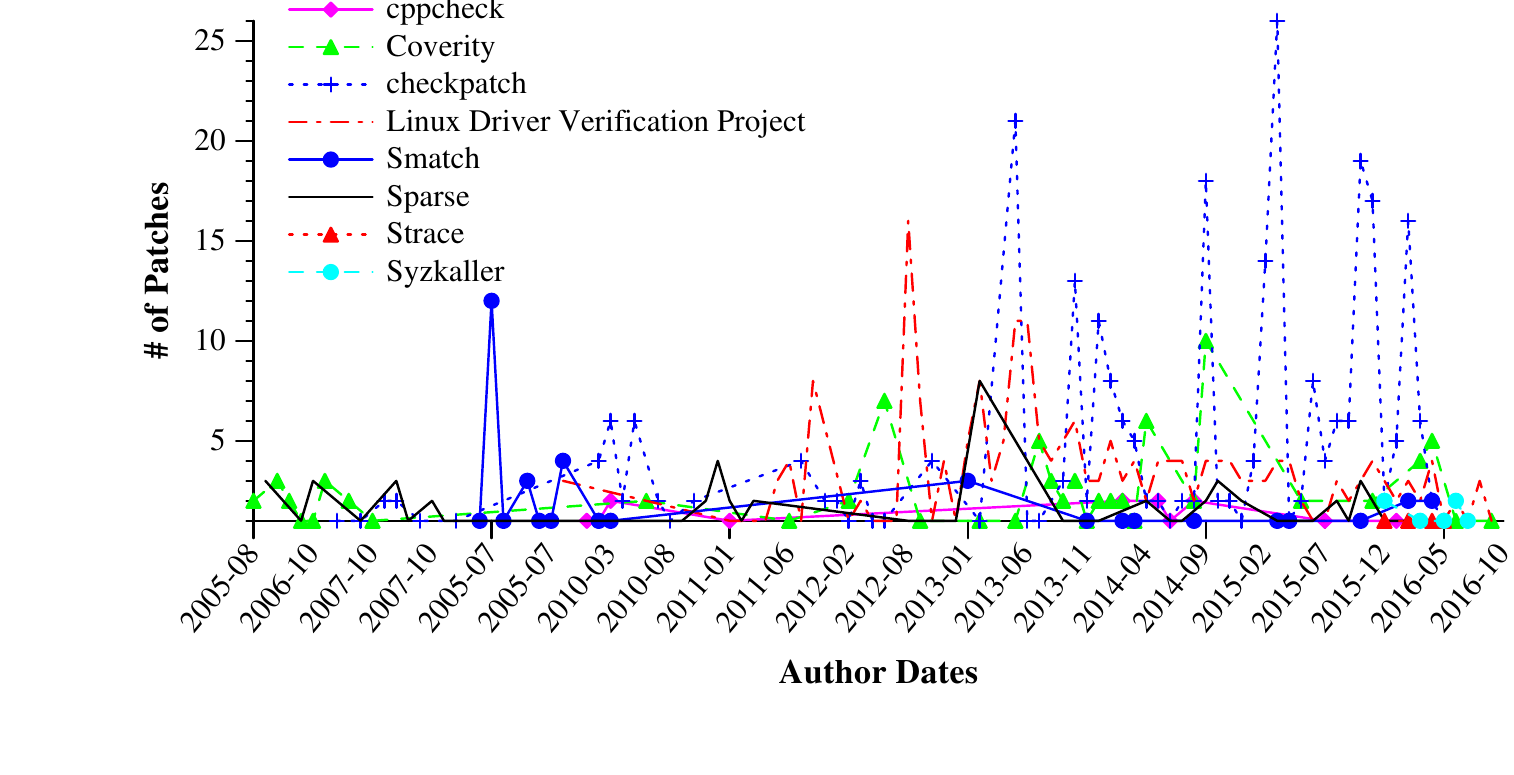}
	\caption{Temporal distributions of DLH patches broken down by tool.}
	\label{fig:temporalSemimanual}
\end{figure}

\begin{figure*}[t]
	\centering
	\subfloat[Number of patches committed by each patch process to Linux's mainline code base.]
	{\parbox{0.38\linewidth}{%
			\includegraphics[width=\linewidth]{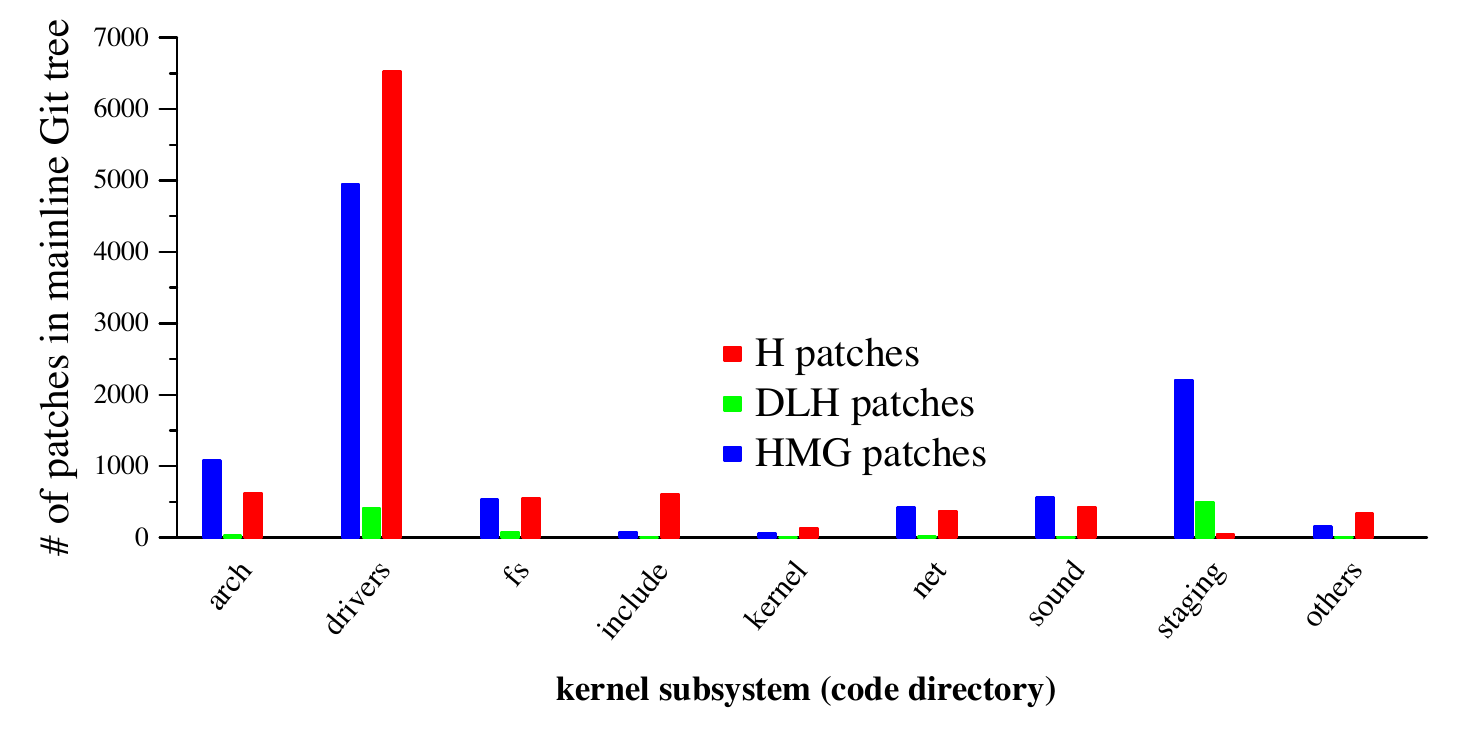}	
	}}%
	\hspace{0.1in}%
	\subfloat[Percentage of patches per subsystem.]
	{\parbox{0.58\linewidth}{%
			\includegraphics[width=\linewidth]{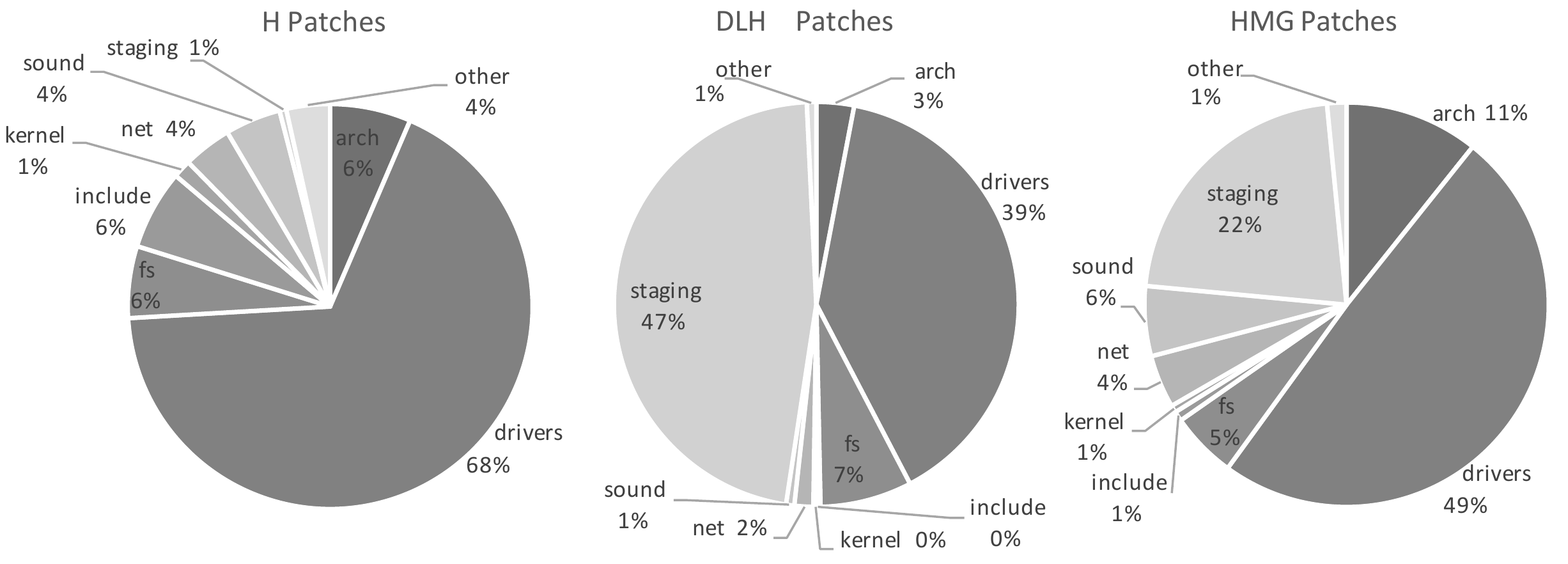}
	}}%
	\caption{Spatial distribution of patches.}
	\label{fig:spatial}
\end{figure*}

Figure~\ref{fig:temporalSemimanual} represents the detailed temporal evolutions
of DLH patches.  Checkpatch, after a slow adoption, is now commonly used, 
followed by Coverity, which regularly contributes to fix vulnerabilities and common
operating system errors. Linux driver verification project tools and Smatch find fewer issues
in mainline code base; such tools are indeed extensively 
used by developers before code is committed in the code base.

%\subsubsection*
{\bf Spatial distribution of patches.}
We compute the spatial distribution of patches across Linux sub-systems.
Linux Kernel's code is split into several folders, each roughly containing all code related to a 
specific sub-system such as file systems, device drivers, architectures, networking, etc. 
We investigate the scenarios of patches with regards to the folders where the files 
are changed and the results are shown in Figure~\ref{fig:spatial}. Most patches are targeted
to device drivers code, and code in early development (i.e., in {\em staging/}\footnote{{\em staging} is a sub-directory of {\em drivers} and contains code that does not yet meet kernel coding standards. We thus separate its statistics from statistics of {\em drivers}.}) that is not yet part of the running kernel. It is noteworthy that header code ({\em include/}), core kernel code 
({\em kernel/}), and to some extent file system code ({\em fs/}), which have been extensively
tested over the years, remain repaired mainly in an all-human process.

Driver code in general, and {\em drivers/staging/} code, in particular, appear to be the place
where tool support is most prevalent. Percentages distribution
in Figure~\ref{fig:spatial}b shows that half (46\%) of DLH patches are targeted at {\em staging} code. 39\% of DLH patches are applied to driver code. Several studies~\cite{Palix:2011:FLT:1950365.1950401,palix_faults_2014,Chou:2001:ESO:502034.502042}
have already shown that driver and staging code contained most kernel errors identified 
by static analysis tools. Similarly, HMG patches are applied in a large majority in drivers code and staging code.

%\vspace{-0.3cm}

\subsection{Acceptance of Patches (RQ1)}
 \label{sec:rq1}

We investigate the reaction of the developer community to the introduction of
bug finding and patch application tools. To that end, we explore, first, the delays in integrating commits, then,
the gaps between the number of patches proposed to the Linux community and those that are finally integrated.

%\subsubsection*
{\bf Delay in commit acceptance.}
Kernel patches are change suggestions proposed by developers to maintainers who often need
time to review them before propagating the changes to the mainline code base. Thus, depending on several factors ---
including the criticality of the bug, complexity of the fix, reliability of the suggested fix, and patch quality ---
there can be a more or less significant delay in commits. 

\begin{figure}[!ht]
	\includegraphics[width=1\linewidth]{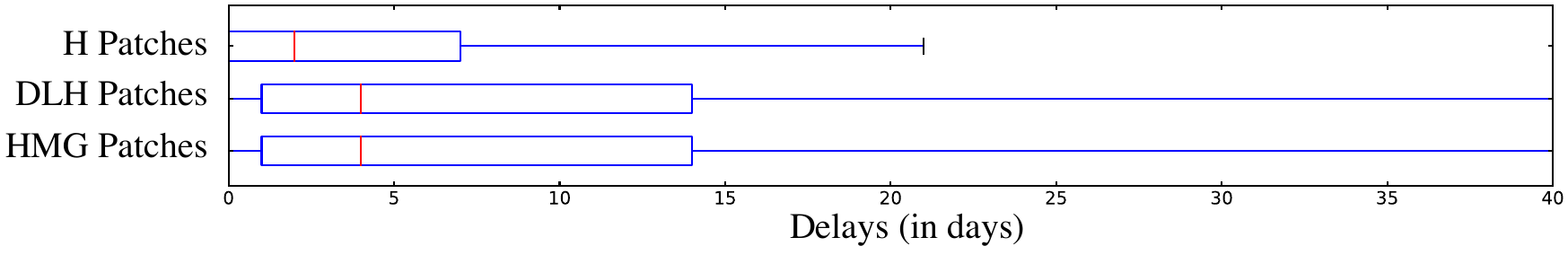}
	\caption{Delay in commit acceptance.}
	\label{fig:delay}
\end{figure}

We compute a delay in commit acceptance as the time
difference between the author contribution date and the commit date (i.e., when the maintainer propagated the
patch to mainline tree). 
Figure~\ref{fig:delay} shows the distribution of delays in the
three different patch construction processes. Overall, H patches appear to be 
more\footnote{We have checked with the Mann-Whitney Wilcoxon test that the difference between 
% \dongsun{we don't need to specify ``median'' here.}	
%median 
delay values is statistically significant.} 
rapidly propagated (median = 2 days) than DLH (median = 4 days) and HMG patches (median = 4 days).  

%\subsection{Tool reference in the Linux Kernel Mailing List (LKML)}
%

%\subsubsection*
{\bf Gaps between discussion and acceptance trends.}
A patch represents the conclusion of an email exchange between the patch author and the 
relevant maintainers about the correctness of the proposed change. As the discussion takes 
place in natural language, it is difficult to categorize how the use of bug finding and patching tools
are valued in the process. Nevertheless, we can use the mailing list to study the frequency at which
developers specifically mention bug finding tools when a patch is first submitted.
Then, we can correlate this frequency on a monthly basis with the corresponding statistics on
accepted DLH patches related to the specific tools.

\begin{figure}[!ht]
	\centering
	\subfloat[Data on checkpatch-related (DLH) patches.]
	{\parbox{0.65\linewidth}{%
			\includegraphics[width=\linewidth]{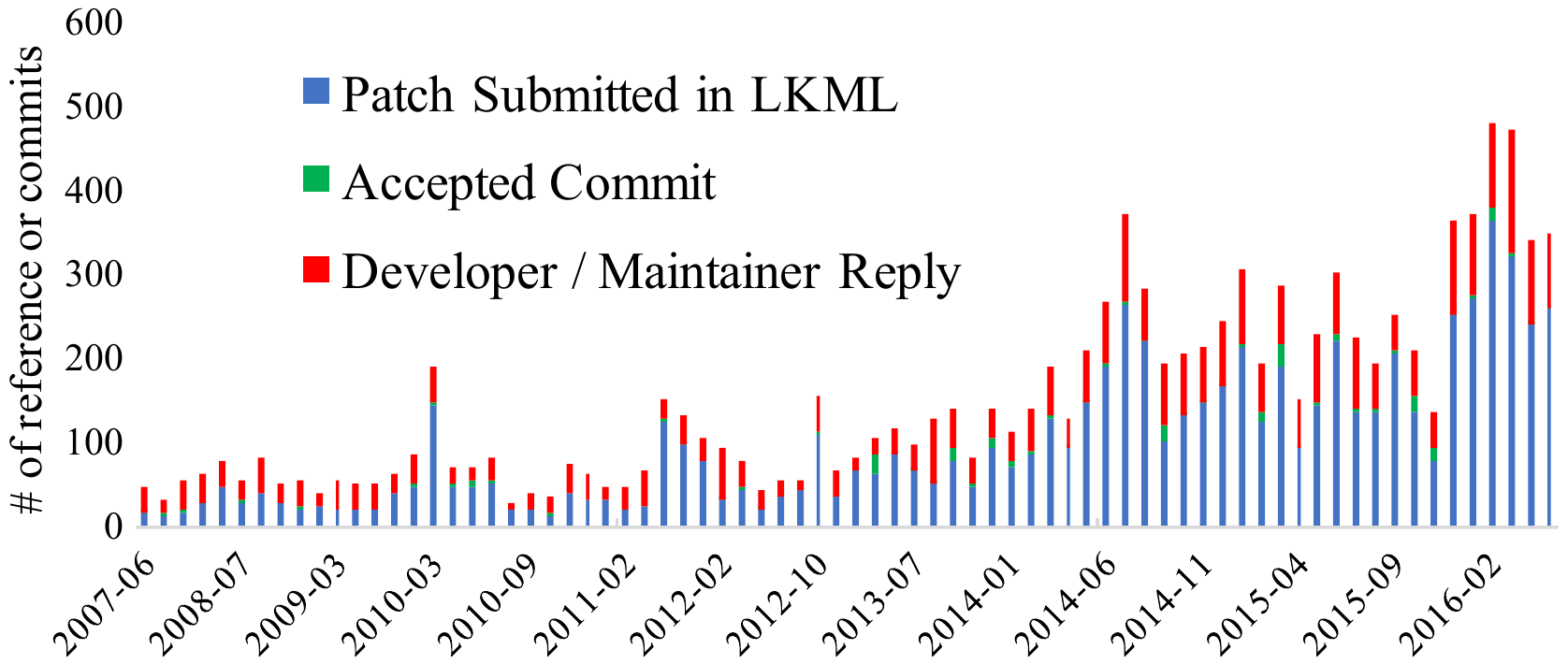}	
	}}%
	\hspace{0.01in}%
	\subfloat[Evolution of the Gap.]
	{\parbox{0.34\linewidth}{%
			\includegraphics[width=\linewidth]{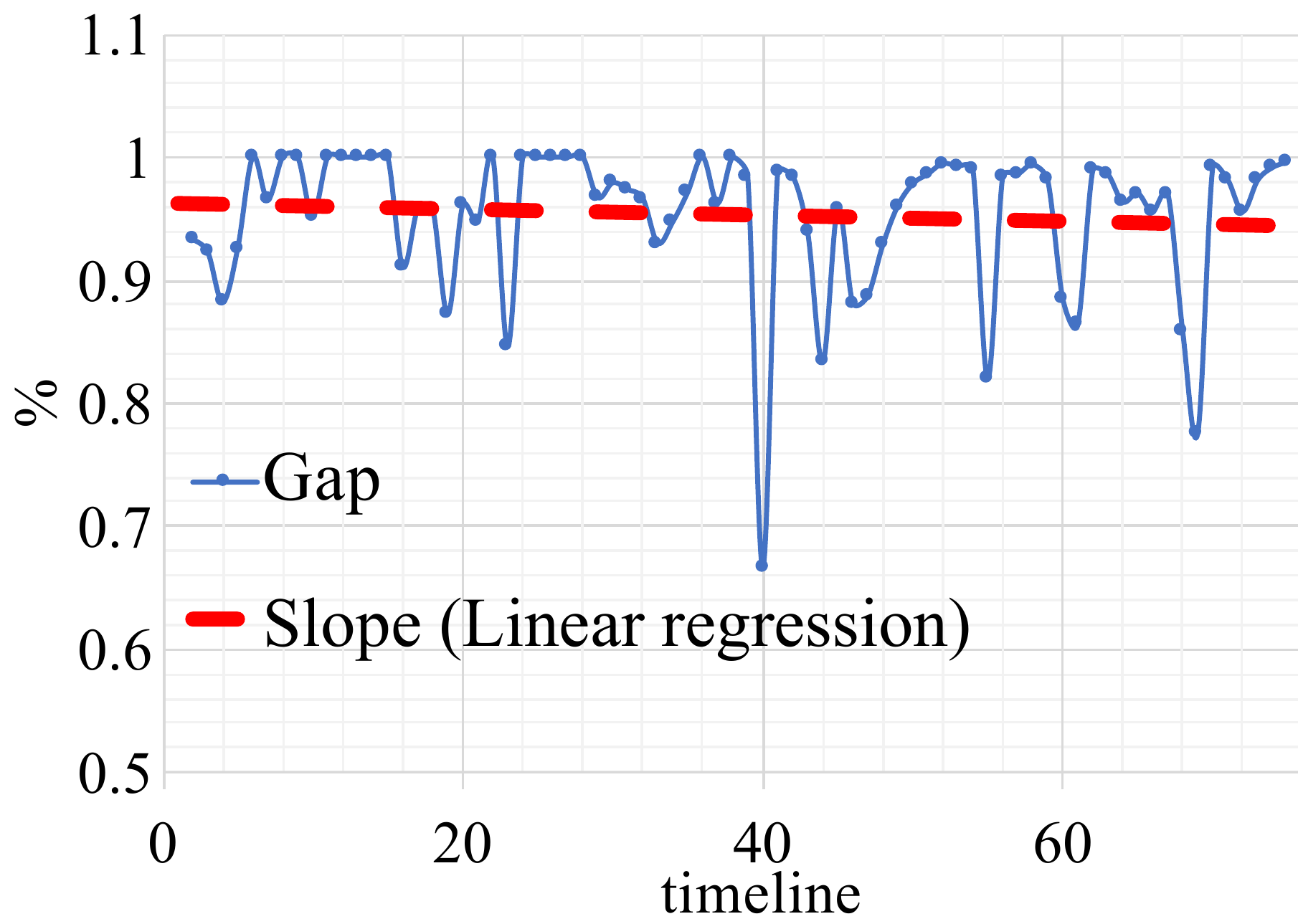}
	}}%
	\hspace{0.1in}%
	\subfloat[Data on coccinelle-related (HMG) patches.]
	{\parbox{0.65\linewidth}{%
			\includegraphics[width=\linewidth]{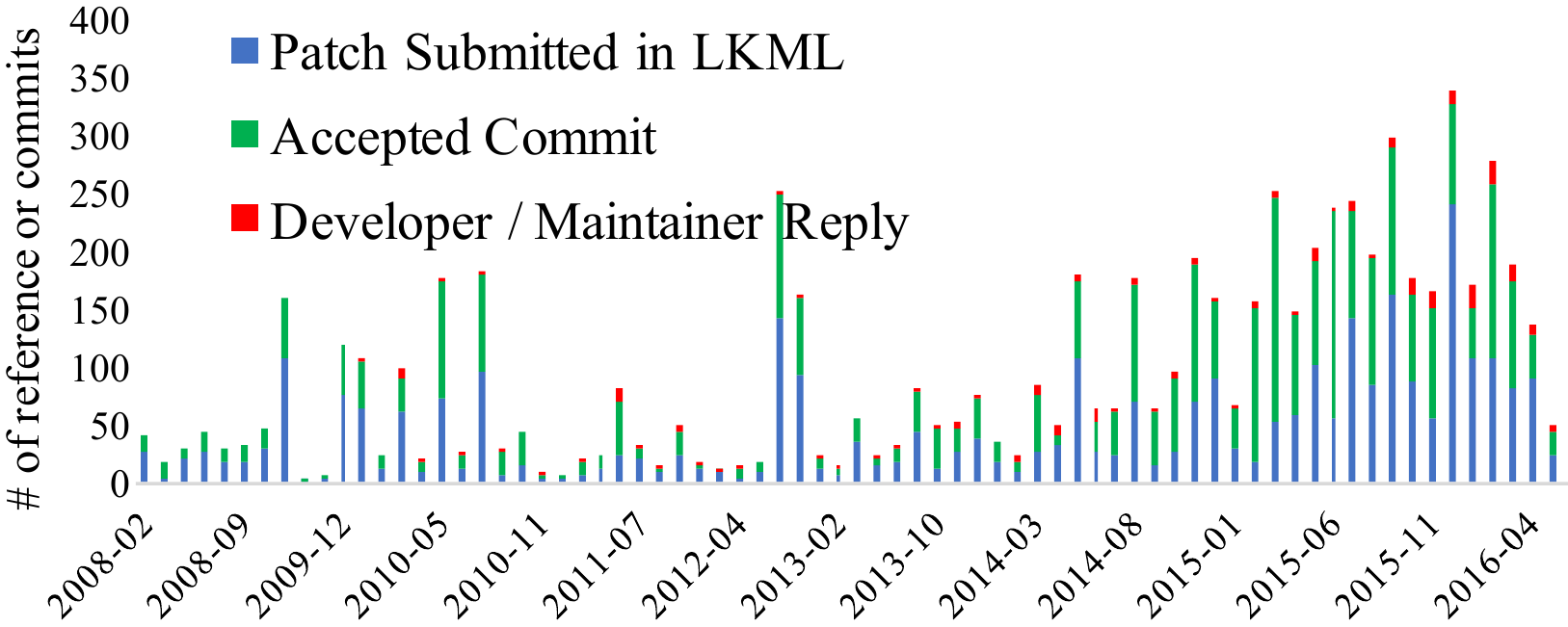}
	}}%
	\hspace{0.01in}%
	\subfloat[Evolution of the Gap.]
	{\parbox{0.34\linewidth}{%
			\includegraphics[width=\linewidth]{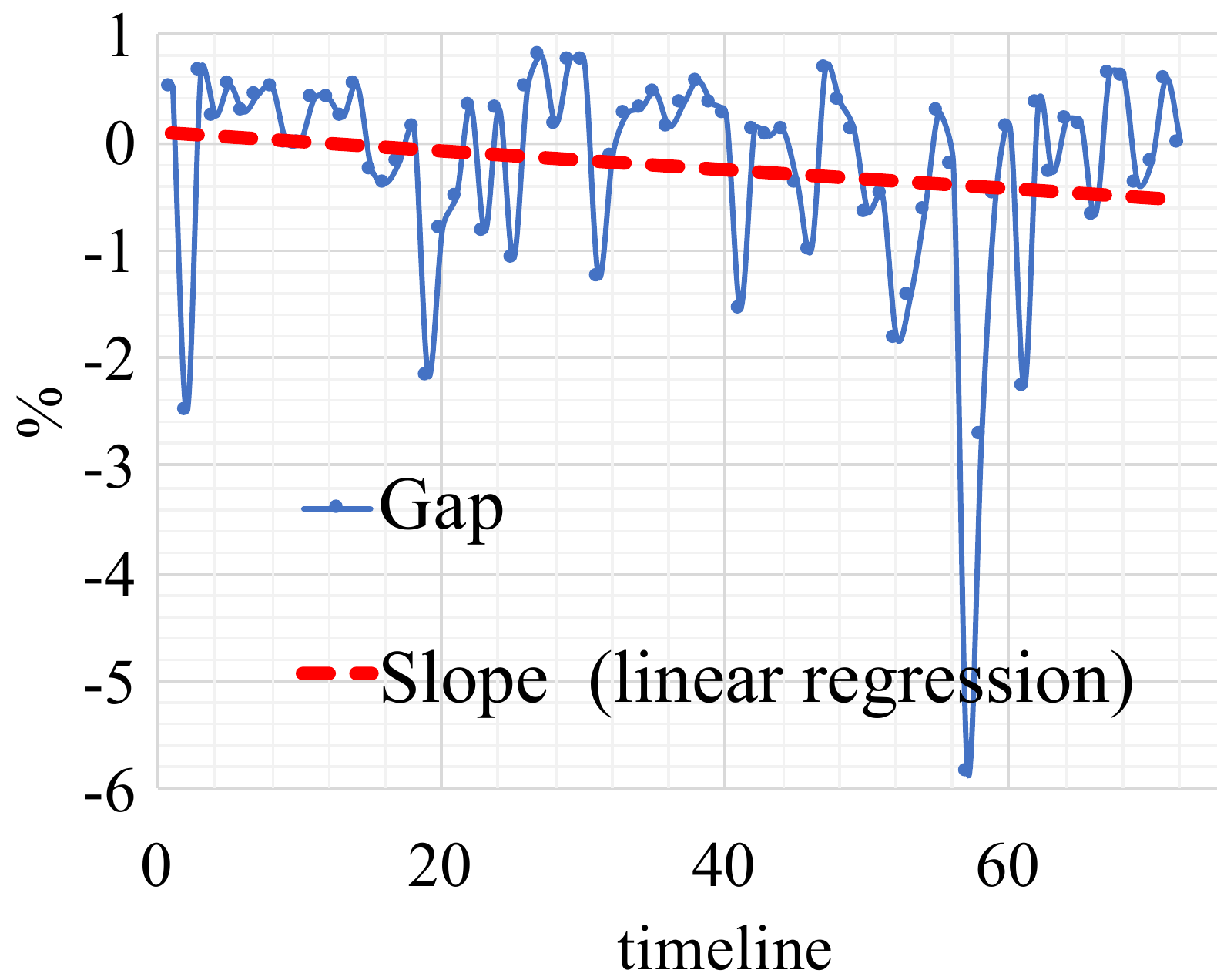}
	}}%
	\caption{\# of Patches submitted / discussed / accepted.}
	\label{fig:lkml}
\end{figure}

We have crawled all emails archived in the Linux Kernel Mailing List (LKML) using Scrapy\footnote{\url{https://scrapy.org/},
a framework for deploying and running spiders}. We use heuristics to differentiate message replies from original
mail content: we consider lines starting with `>' as part of a previous conversation. Finally, we naively search for the
tool name reference in the message text. In total, we crawled\footnote{7,510 entries were empty messages and were thus dropped out.} 1,601,606 original email messages and 885,814 reply messages. As examples, we provide in Figures~\ref{fig:lkml}a and \ref{fig:lkml}c  the distributions per month of the number of patches that were submitted through LKLM mentioning {\em checkpatch} or {coccinelle} respectively, as well as the number of maintainer replies referencing those tools, and the number of related commits accepted into the mainline git tree. To ease observation, we compute in Figures~\ref{fig:lkml}b and \ref{fig:lkml}d the integration gap as a percentage between the number of patches submitted to LKML and the number of patches that are eventually integrated. We draw the slope of the evolution of this gap over time. While checkpatch presents roughly the same gap, the gap is clearly reducing for coccinelle. We have computed the slope for the different sets of tool-supported patches and checked that it was negative for 3 out of 4 of the tools\footnote{We considered only tools associated to at least 50 patches.}: the gap is thus closing over time for most tool-supported processes.

% \begin{table}[!htb]
%\centering
%\caption{Statistics on LKML. \dongsun{hard to interpret. please revise this table.}}
%\resizebox{\linewidth}{!}{
%\label{tab:lkml}
%\begin{tabular}{ l c c c c c }
%\hline
%Patches & Patch Replys & Patches Mentioning Tools & Failed/Empty \\
%\toprule
%
% {\#: 1601606} & {\#: 885814} & {\#: 50421}  & {\#: 7510} \\
% Total: &{\#: 2735535} \\
%\bottomrule
%\end{tabular}
%}
%\end{table}

\begin{tcolorbox}
Tool-supported patches (DLH and HMG alike) have been overall 
accepted at an increasing rate by Linux developers. Integration of such patches by maintainers remains, however, slower than that of traditional H patches.  
\end{tcolorbox}

\subsection{Profile of Patch Authors (RQ2)}
 \label{sec:rq2}

We investigate the speciality and commitment of developers who rely on patch application and
bug finding tools to construct patches.

\begin{figure}[!htb]
	\includegraphics[width=1\linewidth]{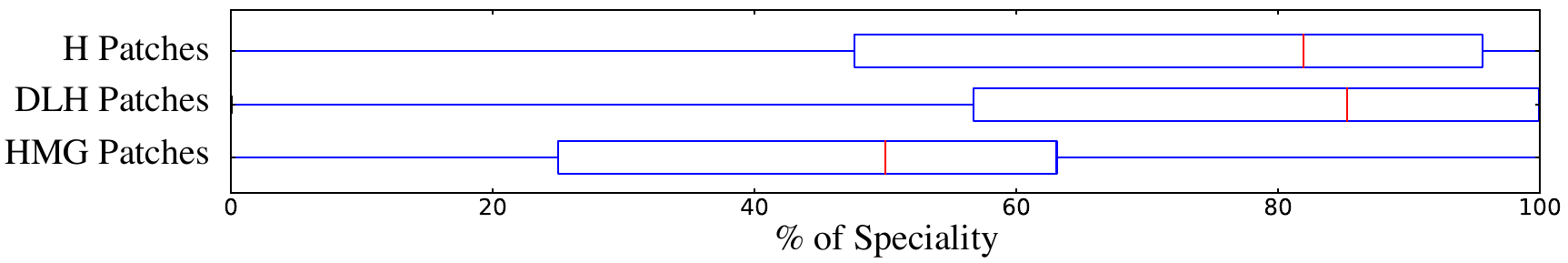}
	\caption{Speciality of developers \texttt{Vs}. Patch types.}
	\label{fig:specialityOfDevelopers}
\end{figure}

{\em Speciality} is defined as a metric for characterizing the extent to which a developer is focused
on a specific subsystem. We compute it as the percentage of patches, among all her/his patches, which a developer
contributes to a specific subsystem. Thus, {\em speciality} is measured with respect to each Linux code directory.
We then draw, in Figure~\ref{fig:specialityOfDevelopers}, the distributions of speciality metric values of developers for the different types of patches: e.g., for
an automated patch applied to a file in a subsystem, we consider the commit author speciality w.r.t that subsystem. %\jk{Maybe difficult to understand. }

H patches are mostly provided by specialized developers. This may imply that the developers focus on implementing specific functionalities over time.
Similarly, DLH patches appear to be mostly applied by specialized developers (even slightly more specialized than those who made H patches). This finding is inline with the requirements
for developers to be aware of the idiosyncrasies of the programming of a particular subsystem to validate the warnings
of bug detection tools and sift through various false positives to produce patches that are eventually accepted by maintainers. HMG patches, on the other hand, are performed by developers on subsystem code which they are not known to be specialized on.

To measure developer {\em commitment}, we follow the approach of Palix et al.~\cite{palix_faults_2014}
and compute, for each developer, the product of (1) the number of patches (H, DLH or HMG)  that have been integrated into Linux and (2) the number of days between the first patch and the last patch.
This metric favours both developers who have contributed many patches over a short period of time and developers who have contributed fewer patches over a longer period of time: e.g., a developer who gets 10 commits
integrated during one year, will have the same degree of commitment as another developer who
gets 40 commits integrated in 3 months.

\begin{figure}[!htb]
	\includegraphics[width=1\linewidth]{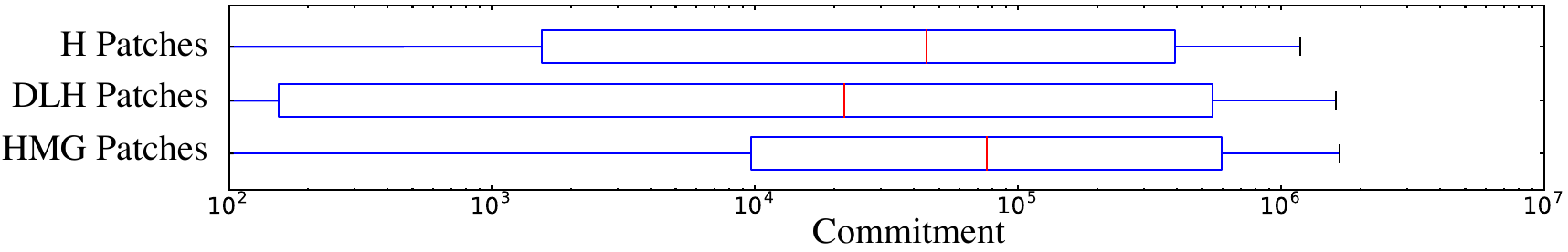}
	\caption{Commitment of developers \texttt{Vs}. Patch types. }
	\label{fig:expertiseOfDevelopers}
\end{figure}

Developer {\em commitment} is studied here as an approximation of developer expertise,
since the more a developer works on the Linux project or with a tool, the more expertise
the developer may be assumed to acquire (on the Linux project and/or with the use of the tool).
Figure~\ref{fig:expertiseOfDevelopers} shows the distribution
of commitment scores of developers for the different types of patches.

DLH patches are shown to be produced by developers with a more varying degree of
commitment (greater standard deviation). The median value of commitment is further lower than
the median commitment for HMG patches. Finally, overall, the distributions of commitment values
of developers indicate that H patch authors present lesser commitment than HMG patch authors.

We then use Spearman's $\rho$~\cite{Spearman} to measure the degree of correlation between the commitment of developers and the number of tool-supported patches that they submit. We focus on specialized\footnote{speciality metric value greater than 50\%} developers of two very different kinds of code: mature file system ({\em fs}) code and early-development ({\em staging}) code. The correlation is then revealed to be higher ($\rho=0.42$) for staging than for fs ($\rho=0.11$). We also note that 64\% of developers committing code in staging stick to this part of the code for over half of their contributions. Finally, developers specialized in {\em kernel} have never relied on tool support to produce a patch.

\begin{tcolorbox}
Bug detection tools are generally used by developers with (to some extent)  knowledge of the code. Patch application tools, on the other hand, enable developers to remain committed to contributing patches to the code base.
\end{tcolorbox}

\subsection{Stability of Patches (RQ3)}
 \label{sec:rq3}
Although patches are carefully validated before they are integrated to the mainline code base, a patch might be simply incorrect and thus the relevant code may require further changes, or the patch may simply be reverted. 
However, it is challenging to precisely detect and resolve such a change in recently patched code hunks. Even this requires heuristics that may prove to be error-prone. 
Thus, in this study, we focus on commits whose reverting is explicit. 

It is common for software developers to cancel patches that they hastily committed to the code base. The {\tt git revert} command is an excellent means for developers to roll back their commits. However, given the hierarchical organization in Linux, when a patch has reached the mainline, a simple revert (using git commands) is uncommon. The submitting developer (or another one) must write another patch explaining the need to revert. This patch again goes through the process to be accepted in the mainline. In this setting, the revert of a commit is likely strongly justified. We search for commits that are reverted by looking at commit messages where we have seen a pattern of the form ``{\tt revert <hash>}''\footnote{We use: \texttt{git show '+sha+' | grep -E -i "revert .[0-9a-f]{5}+ | commit .[0-9a-f]{5}+ | [0-9a-f]\{40\}\$*}}. 
		
\begin{figure}[!htb]
	\includegraphics[width=1\linewidth]{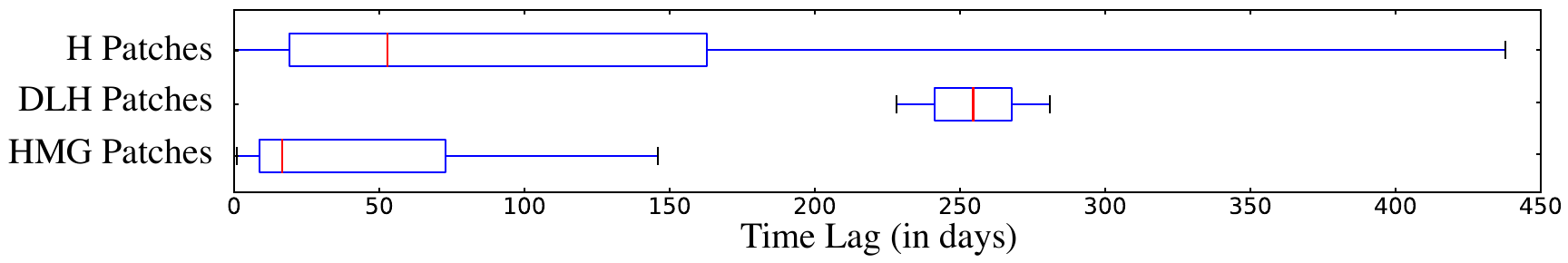}
	\caption{Time lag between patch integration and reverting.}
	\label{fig:revertedCommits}
\end{figure}

We have found that 2.81\% of H-patch commits have been later reverted. In contrast, only 0.27\% and 0.32\% respectively of DLH and HMG patch commits have been reverted. Figure~\ref{fig:revertedCommits} further provides the distributions of delays in reverting commits.
	
H-patches revert delay distribution is the most spread. On average (median), a DLH patch, when it is reverted, will be so after 250 days (8 months). On the other hand, HMG patches 
will be reverted in less than a month (20 days). The median delay for revert is of 60 days for H patches.

\begin{tcolorbox}
Tool-supported patches are generally stable. However, while patches fixing tool warnings may be found inadequate long after their integration, issues with patches generated based on fix patterns appear to be discovered quickly.
\end{tcolorbox}

 \subsection{Bug Kinds (RQ4)}
 \label{sec:rq4}
 We study bug kinds in two dimensions: the spread of buggy code and the complexity of the bugs. We investigate the locality of patches as an approximation of the spread of buggy code, and the change operations at the level of Abstract syntax tree nodes modifications to approximate complexity of bugs.
 
 \subsubsection{Locality of Patches}

 The locality of patches is a key dimension
 for characterizing patches. Patch size has been measured in the 
 literature~\cite{palix_faults_2014,Bissyande:2013:EEB:2495256.2495765} in terms of the number of
 code locations that it involves, while several state-of-the-art automated repair approaches mostly focus on single/limited
 code changes to fix software. The Linux project is a particularly adequate study subject
  for this comparison since developers are often reminded that they must ``solve a single problem per patch''\footnote{see {\tt Documentation/SubmittingPatches}}: fix operations are then generally separated from cosmetic changes.
 
\begin{figure}[!htb]
\begin{center}
\includegraphics[width=\linewidth]{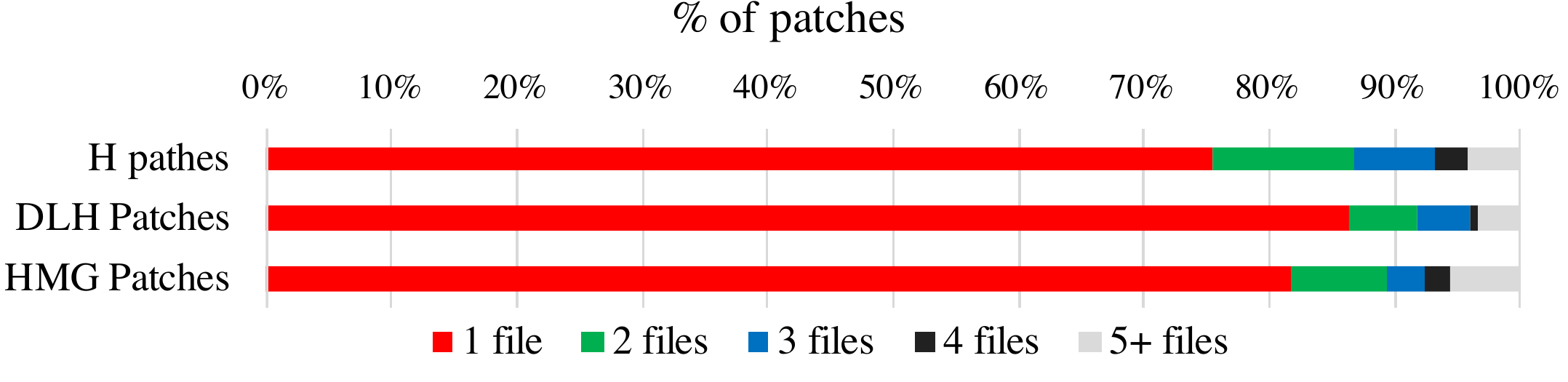}
\caption{Distribution of patch sizes in terms of files.}
\label{fig:locality-files}
\end{center}
\end{figure}

 A bug fix patch may involve changes across files. Figure~\ref{fig:locality-files}
shows that most fixes are localized to a single file independently of the way they
are constructed. 

DLH patches appear to
be the more local, while more than 20\% of H patches implement simultaneous changes
in at least two files. Interestingly, we note that HMG patches include the largest
proportion of patches (5.6\%) that simultaneously change 5 files or more. Such patches are
generated to fix pervasive bugs such as the wrong usage of an API, or to implement
a collateral evolution.

\begin{figure}[!htb]
\centering
\includegraphics[width=\linewidth]{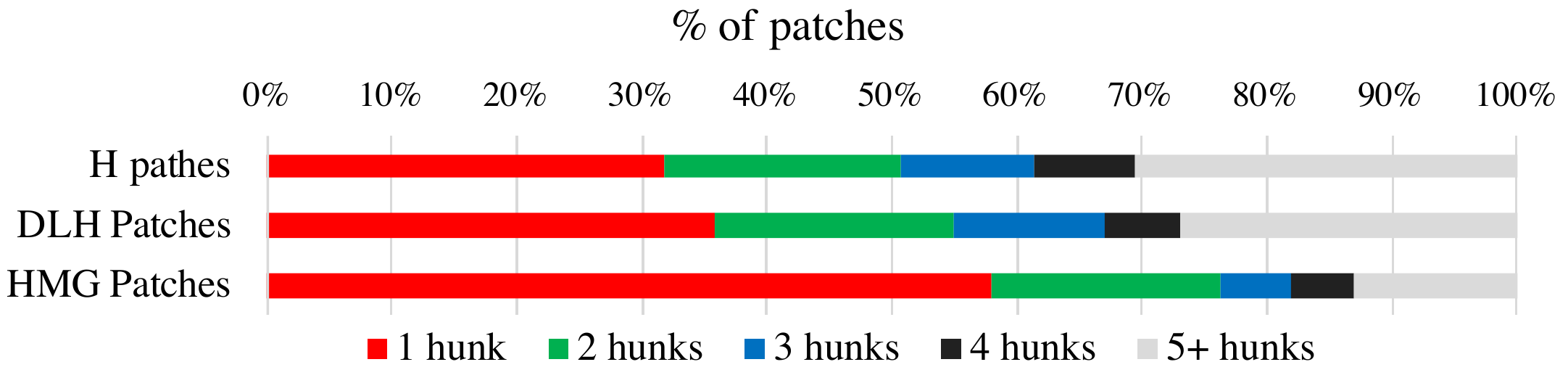}
\caption{Distribution of patch sizes in terms of hunks. }
\label{fig:locality-hunks}
\end{figure}

We further investigate the locality of patches in terms of the number of code hunks (i.e., a contiguous group of code lines\footnote{\url{https://www.gnu.org/software/diffutils/manual/html_node/Hunks.html}})
that are changed by a patch. Indeed, code files can be large, and a patch may
variably spread changes inside the file, which, to some extent, may represent a
 degree of complexity of the fix. Figure~\ref{fig:locality-hunks} shows
that H patches are more likely to involve several hunks of code than HMG 
and DLH patches. 

Our observations on patch sizes suggest that developers, with or without 
bug finding tools, must correlate data and code statements across different code blocks to 
repair programs.  %\dongsun{we need to describe more about this one. Present our interpretation of this data and list up potential implications.}

%\dongsun{Figures 10 and 11 should be redrawn (should rearrange bars and legends).}

\begin{figure}[!htb]
\centering
\includegraphics[width=\linewidth]{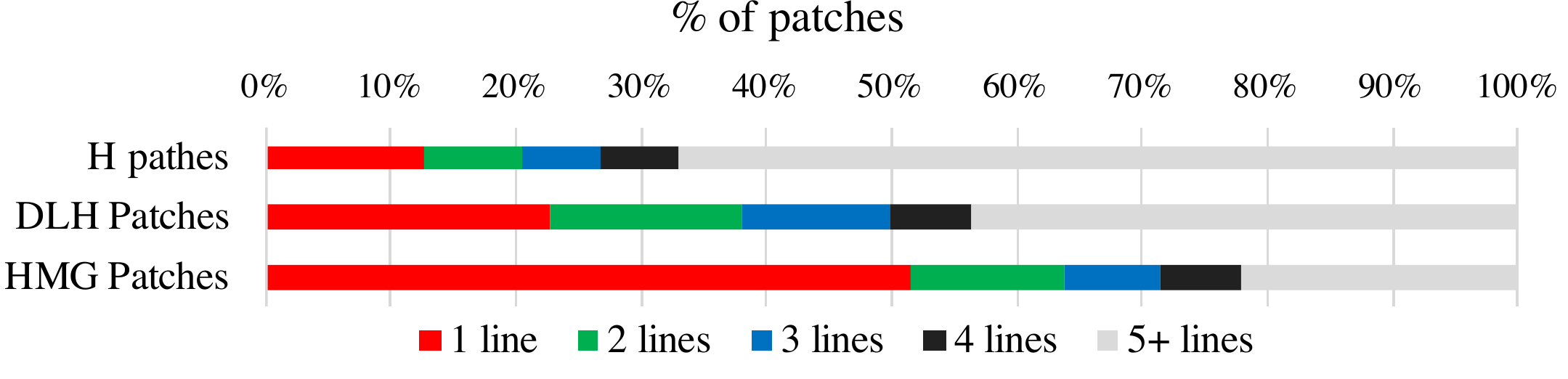}
\caption{Distribution of patch sizes in terms of lines.}
\label{fig:locality-lines}
\end{figure}

Finally, we compute the locality of the patches in terms of the number of lines that 
are affected by the changes. Such a study is relevant for estimating the proportions
of isolated change (i.e., single-line changes) that fix bugs in the three scenarios of repairs.
Figure~\ref{fig:locality-lines} reveals that the large majority of patches that are manually crafted as responses to bug reports change several lines, with almost 70\% patches impacting at least 5 lines. On the other hand,
over 40\% HMG patches impact only at most two lines of code. 

%\tb{@Anil, we can study and see if there is a correlation between severity of patches and the number of patch lines for manual repair}

\begin{figure*}[t]
	\centering
	{\parbox{0.24\linewidth}{%
			\includegraphics[width=\linewidth]{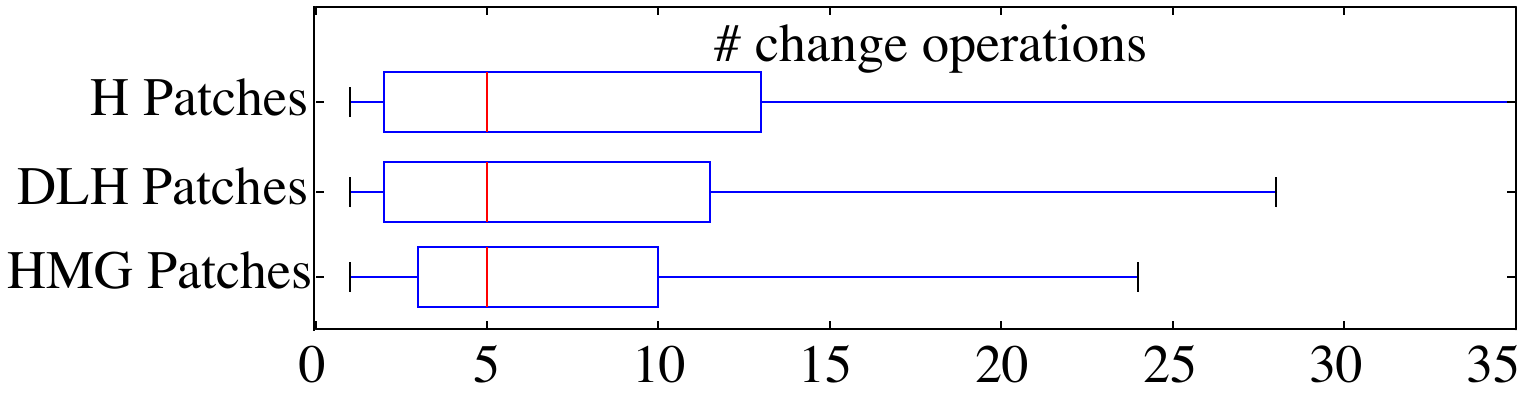}	
	}}%
	\hspace{0.01in}%
	{\parbox{0.24\linewidth}{%
			\includegraphics[width=\linewidth]{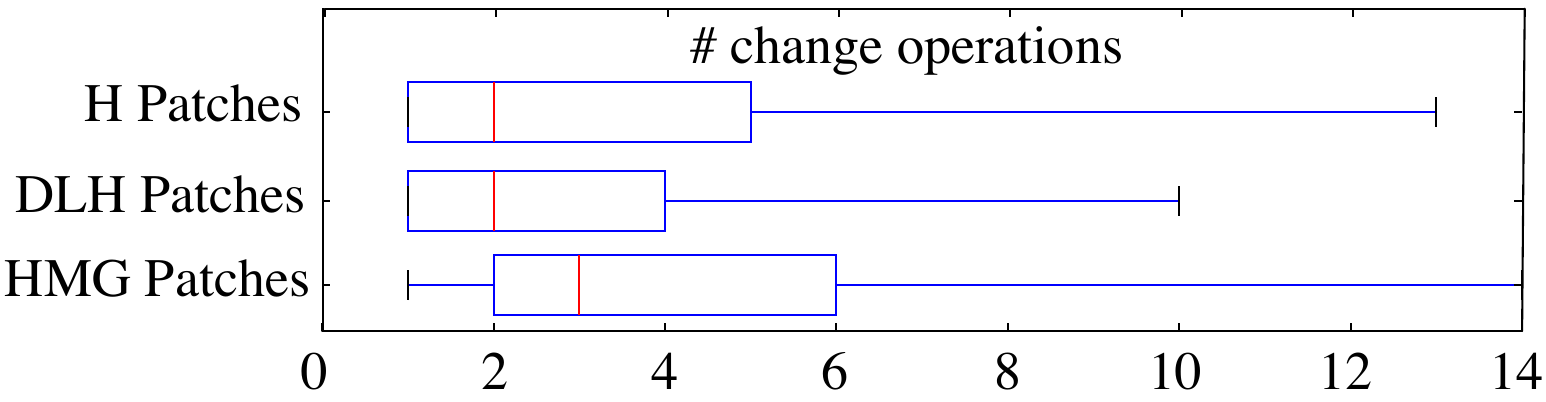}
	}}%
	\hspace{0.1in}%
	{\parbox{0.24\linewidth}{%
			\includegraphics[width=\linewidth]{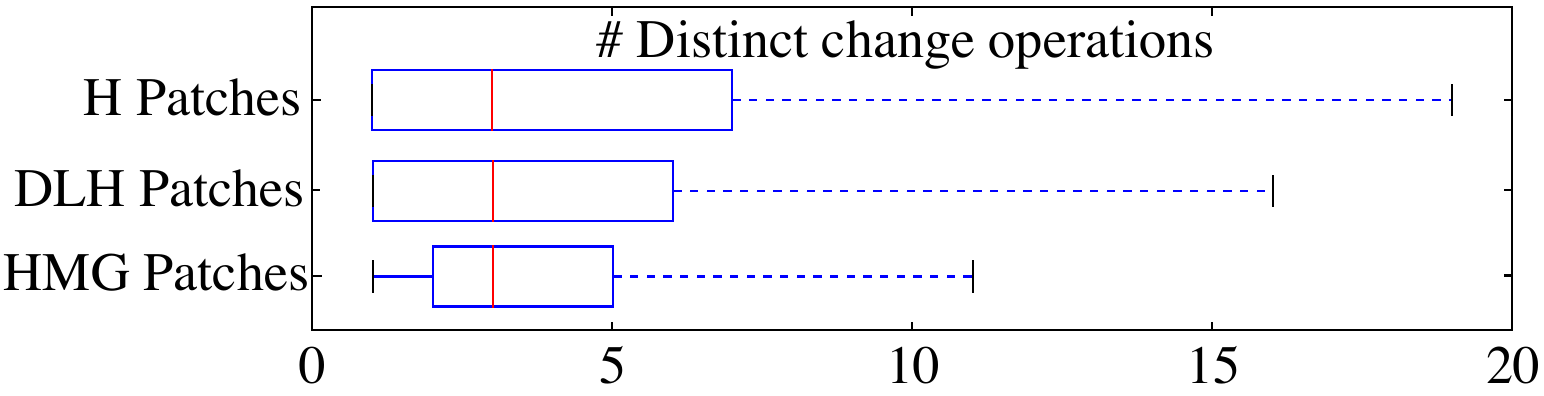}
	}}%
	\hspace{0.01in}%
	{\parbox{0.24\linewidth}{%
			\includegraphics[width=\linewidth]{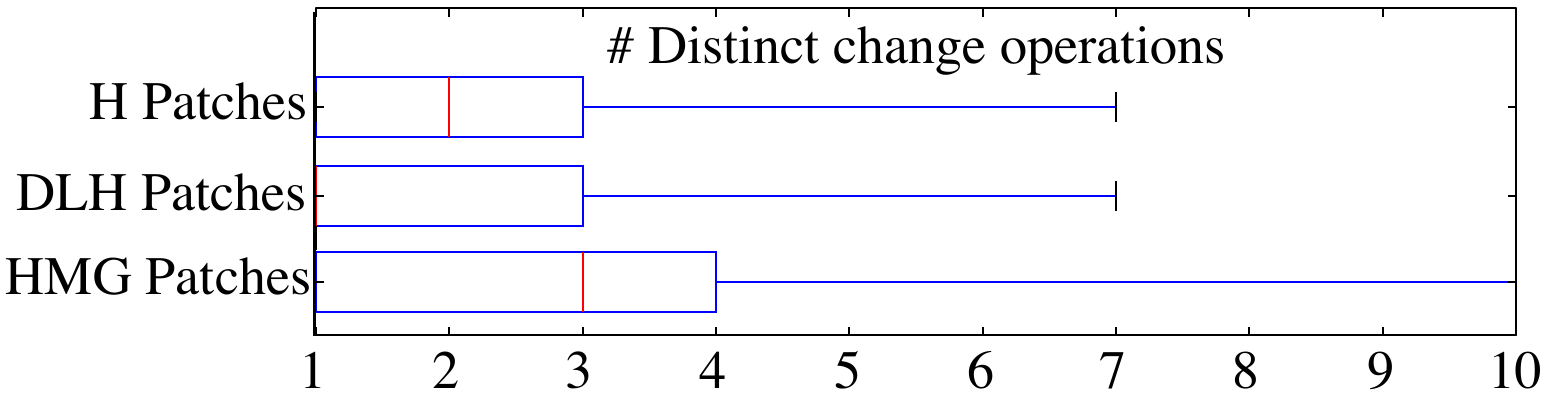}
	}}%
	
	\subfloat[\scriptsize \# of operations / single file.]
	{\parbox{0.24\linewidth}{%
			\includegraphics[width=\linewidth]{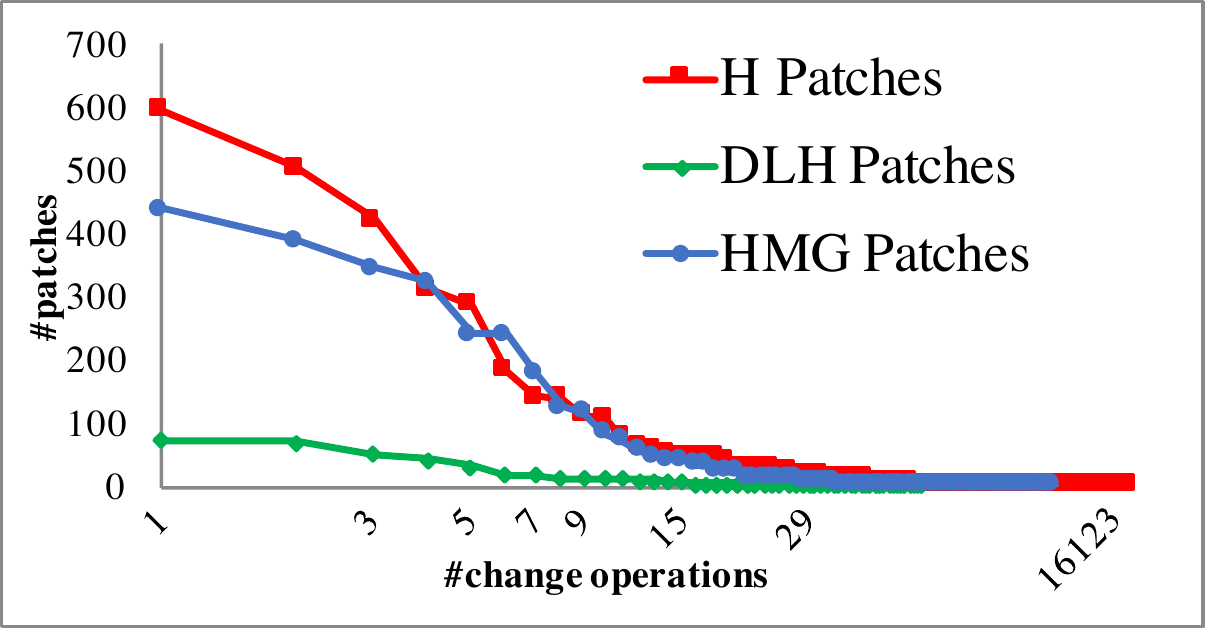}	
	}}%
	\hspace{0.01in}%
	\subfloat[\scriptsize \# of operations / single hunk.]
	{\parbox{0.24\linewidth}{%
			\includegraphics[width=\linewidth]{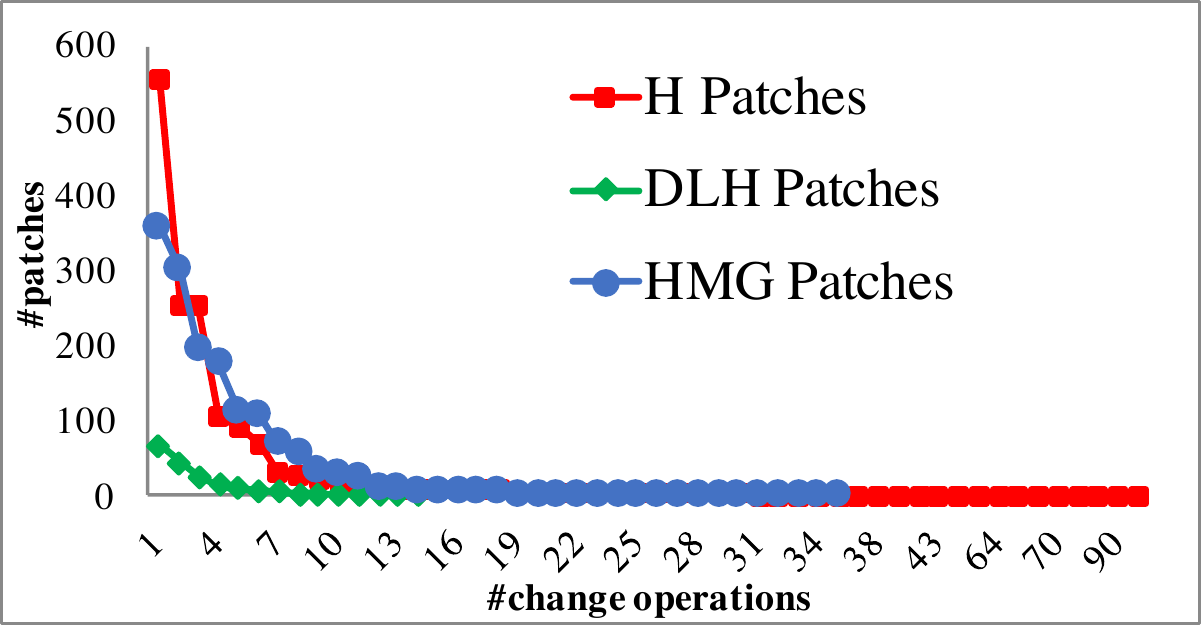}
	}}%
	\hspace{0.1in}%
	\subfloat[\scriptsize \# of distinct operations / single file.]
	{\parbox{0.24\linewidth}{%
			\includegraphics[width=\linewidth]{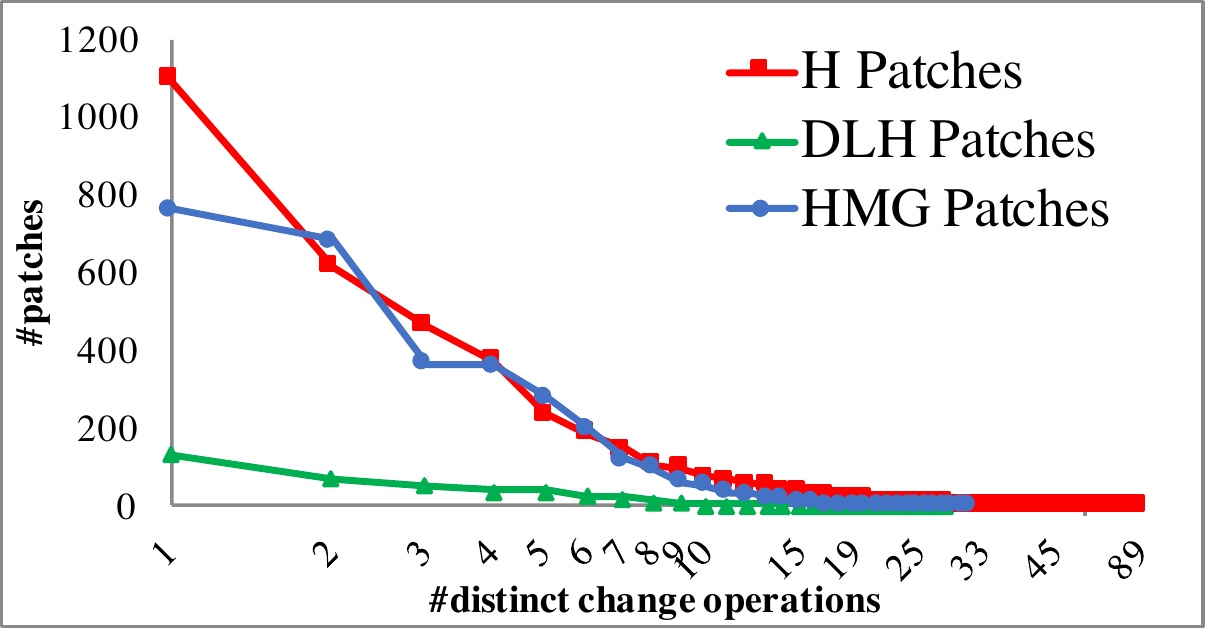}
	}}%
	\hspace{0.01in}%
	\subfloat[\scriptsize \# of distinct operations / single hunk.]
	{\parbox{0.24\linewidth}{%
			\includegraphics[width=\linewidth]{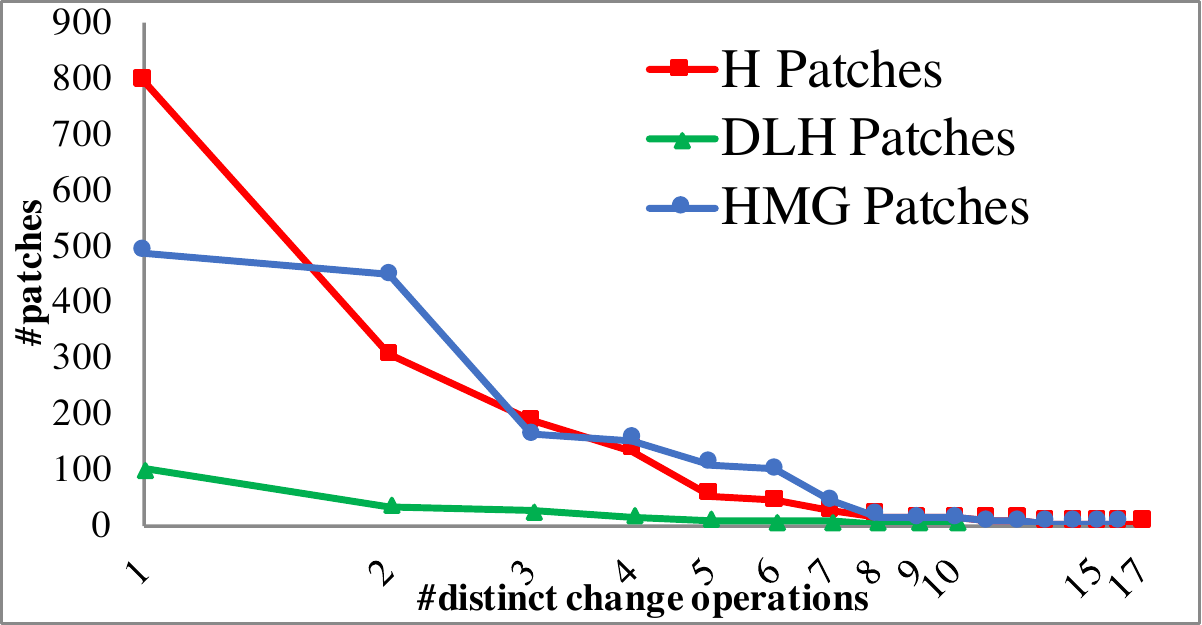}
	}}%
	\caption{Distribution of change operations (Total \# of operations \& \# of distinct operations in patches).}
	\label{fig:operations}
\end{figure*}

\begin{figure*}[t]
	\centering
	\subfloat[H patches.]
	{\parbox{0.33\linewidth}{%
			\includegraphics[width=\linewidth]{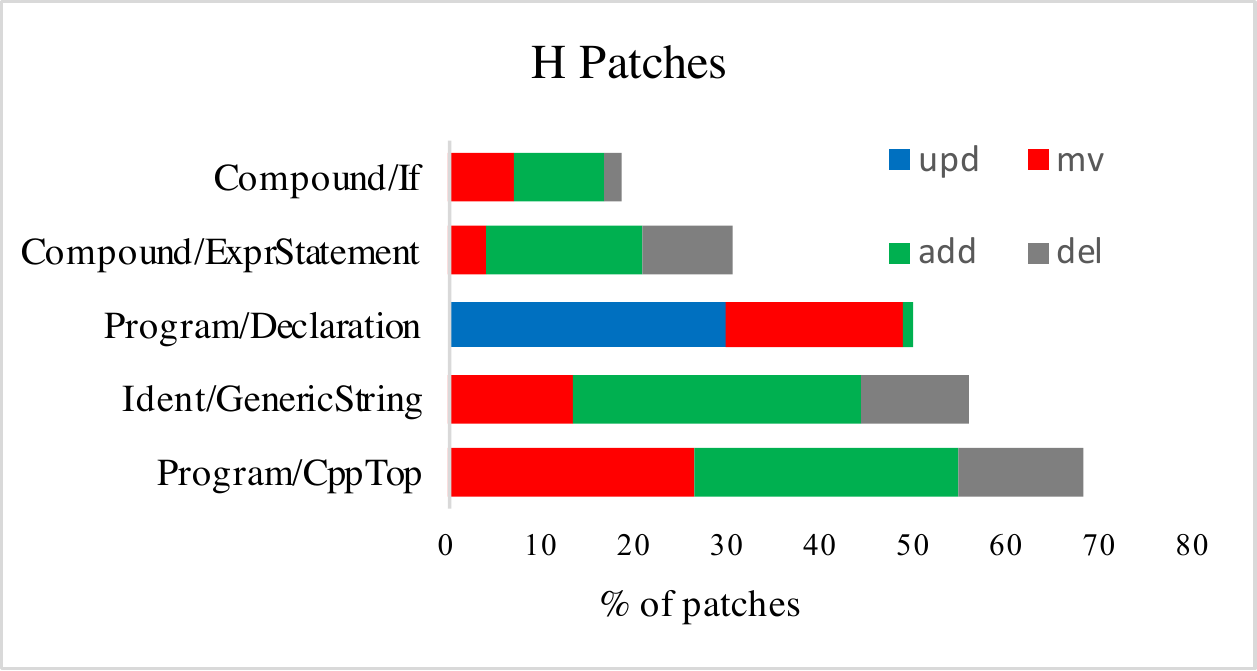}
	}}%
	\hspace{0.01in}%
	\subfloat[DLH patches.]
	{\parbox{0.33\linewidth}{%
			\includegraphics[width=\linewidth]{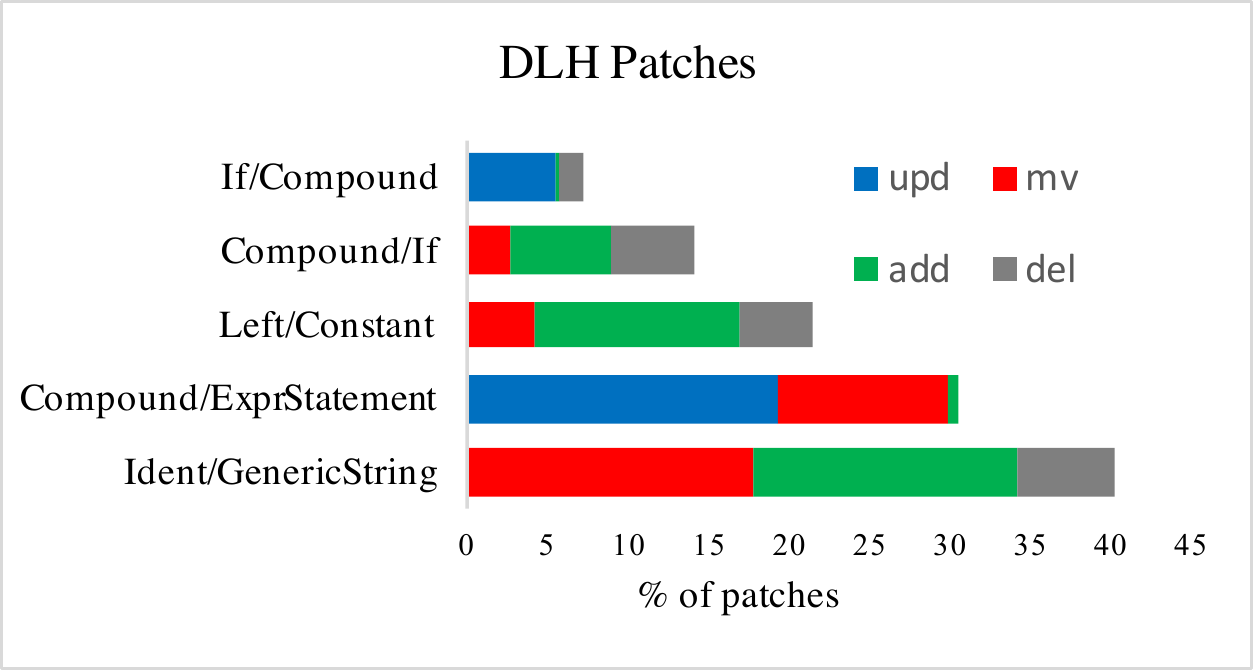}
	}}%
	\hspace{0.01in}%
	\subfloat[HMG patches.]
	{\parbox{0.33\linewidth}{%
			\includegraphics[width=\linewidth]{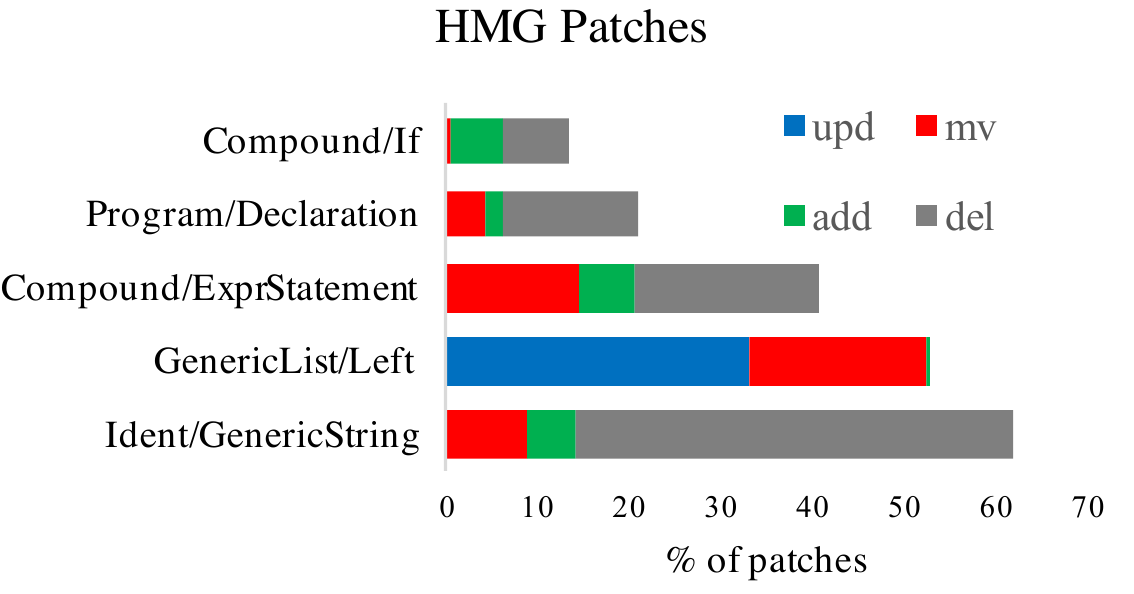}	
	}}%

	\caption{Top-5 change operations appearing at least once in a patch from the three processes.}
	\label{fig:typesOfoperations}
\end{figure*}

\subsubsection{Change Operations in Patches} 
In general, line-based diff tools, such as the GNU Diff,
are limited in the expression of the kinds of changes that can be identified since they consider
only adds and removes, but no moves and updates~\cite{palix_improving_2015}. Thus, to
investigate change operations performed by patches, we rely on approaches that compute
modifications based on abstract syntax trees (AST)~\cite{kim_program_2006}.
Such approaches produce fine-grained results at the level of individual nodes. For this study,
we consider an extended version of the open-source GumTree~\cite{falleri_fine-grained_2014}
with support for the C language~\cite{palix_improving_2015}. This tool specifically takes into account additions,
deletions, updates and moves of individual tree nodes, and
has the goal of producing results that are easier for users to
understand than those of GNU Diff.

The output of GumTree is an edit script enumerating a sequence of operations that must be carried
out on an AST tree to yield the other tree. To that end, GumTree implements a mapping algorithm between
the nodes in two abstract syntax trees. This algorithm is inspired by the way developers manually look
at changes between two files, first searching for the largest unmodified chunks of code (i.e., isomorphic subtrees)
and then identifying modifications (i.e., given two mapped nodes, find descendants that share a large percentage of common mappings, and so on). Given those mappings, GumTree leverages an optimal and quadratic algorithm~\cite{chawathe_change_1996} to compute the edit script. More details on the algorithm can be found in the original articles~\cite{chawathe_change_1996,falleri_fine-grained_2014}.

For simplicity, in this paper, we express change operations in their abstract form as a triplet ``{\em scope/element:action}'' where {\em scope}
represents the type of node (e.g., the program, an {\tt If} block, a compound block, a generic list, an identifier, etc.) where the change occurs, {\em element} represents the element (e.g., an expression, a declaration, a generic string, a compound block, an if block, etc.) that is changed 
and {\em action} represents the move/update/add/delete operators that are used. This abstract representation indeed does not take into account any variable names and functions involved (and available in the output of GumTree).
Figure~\ref{fig:examplesOperations} shows a patch example for a change operation where a new {\ttfamily If} block code is inserted.

\begin{figure}[!htb]
	\centering
%	\subfloat[Update of field access:  {\tt RecordPtAccess/GenericString:upd}]
%	{\parbox{0.49\linewidth}{%
%		\lstinputlisting[linewidth={\linewidth}, frame=tb]{fig/record-genstring.list}
%}}%
%	\hspace{0.1in}%
%	\subfloat[Addition of an If bloc: {\tt Compound/If:add}]
{\parbox{1\linewidth}{
\begin{lstlisting}[linewidth={\linewidth}, frame=tb,basicstyle=\scriptsize\ttfamily]
diff --git a/drivers/gpu/drm/i915/intel_display.c b/.../drm/i915/intel_display.c
index 6e0d828..182f849 100644

--- a/drivers/gpu/drm/i915/intel_display.c
+++ b/drivers/gpu/drm/i915/intel_display.c
@@ -13351,6 +13351,9 @@ int intel_atomic_prepare_commit(struct drm_device *dev,
        for_each_crtc_in_state(state, crtc, crtc_state, i) {
+               if (state->legacy_cursor_update)
+                       continue;
+
                ret = intel_crtc_wait_for_pending_flips(crtc);
\end{lstlisting}
}}%
	\caption{Example of {\tt Compound/If:add} -- Add an {\tt If} block.}
	\label{fig:examplesOperations}
\end{figure}

\begin{figure*}[!htb]
\centering
	\includegraphics[width=\linewidth]{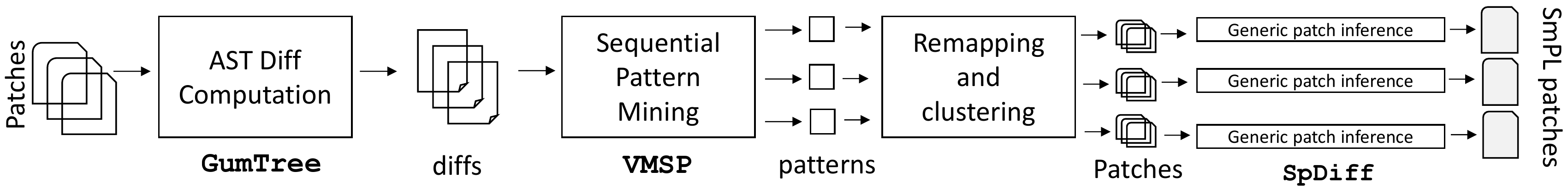}
	\caption{Searching for redundancies among patches that fix warnings of bug finding tools (i.e., DLH patches).}
	\label{fig:mutation-operations}
\end{figure*}

Figure~\ref{fig:operations} illustrates the distributions of the number of operations that are performed in a patch. To limit the bias of changes that are identically performed in several files (e.g., Coccinelle collateral evolutions), we focus on patches that touch a single file, then on patches that are limited to a single hunk. All distributions are long-tail, revealing that most patches apply very few operations in terms of number and variety. 
While the three processes have similar average (median) values of change operations performed on a file, HMG patches appear to implement changes with a consistent number of operations (limited standard deviation). On the other hand, when we consider change operations at the hunk level, DLH patches apply fewer operations than HMG patches\footnote{We have checked with MWW tests that the difference is statistically significant.}.

Figure~\ref{fig:typesOfoperations} summarizes the top-5 change operations that
are recurrently implemented by patches constructed in the different processes considered in our study. Changes performed appear to be specific for each process. For example, while {\tt Ident/GenericString} and {\tt Compound/If}-related change operations occur in most patches, they do not display the same proportions in terms of additions, moves, updates and deletions. %Patches created to fix bug tools warnings 

\begin{tcolorbox}
Overall, patches, following their construction process, differ in terms of size (i.e., the spread of the buggy code that they repair) and in the nature of change operations that they implement (i.e., the complexity of the bug). 
\end{tcolorbox}

%\input{sections/api}

%\begin{figure*}[!htb]
%\centering
%	\includegraphics[width=\linewidth]{jgraph/spdiff_process2}
%	\caption{Searching for redundancies among semi-manual patches.}
%	\label{fig:mutation-operations}
%\end{figure*}

 \section{Discussions}
 We discuss the implications of our findings for the software engineering research community,
 in particular, the automated research field, and enumerate the threats to validity that this study carries.

\subsection{Implications}
As the field of automated repair is getting mature, the community has started to reflect
(i) on whether to build human-acceptable or readable patches~\cite{kim_automatic_2013,monperrus_critical_2014},
(ii) on the suitability of automated repair fixes~\cite{smith_is_2015},
(iii) on the relevance of patches produced by repair tools~\cite{zhong_empirical_2015}.
Our work continues this reflection from the perspective of the acceptance of tool-support
in patch construction.
We further acknowledge that HMG patches considered in this study are not constructed in the same spirit as
in automated repair: indeed, automated repair approaches make no a-priori assumption on what and where
the fault is, while tools such as Coccinelle~\cite{Brunel:2009:FFP:1480881.1480897} produce patches based on fix patterns that match buggy code locations.
Nevertheless, given the lack of integration of automated repair in a real-world development process,
we claim that investigating Linux patch cases can offer insights which can be leveraged by the research
community to understand how the developer community can accept tool-supported patches, and the automation of what kind of fixes can be readily accepted in the community.

%\subsubsection*
{\bf On manual Vs. tool-supported patches.} As illustrated in Section~\ref{sec:stat}, tool-supported patch construction is becoming frequently and widely used in the Linux Kernel development. In particular, HMG patches account for a larger portion of recent program changes than H patches. This suggests that both (1) developers gradually accept to use patch application tools such as Coccinelle~\cite{Brunel:2009:FFP:1480881.1480897} since they are effective to automatically change similar code fragments and (2) there are many (micro) code clones~\cite{tonder_defending_2016} in the code base. Regarding spatial distribution, DLH and HMG patches are committed to `staging' (22-47\%) while H patches in `staging' account for only 1\%. This may indicate that experimental features have more opportunities for tools to help write bug fixing patches.
It implies indeed that, for early development code, the community almost exclusively relies upon tools to solve common bugs (e.g., in relation with programming rules, styles, code hardening, etc.) by novice programmers (i.e., not necessary specialized in kernel code), before expert developers can take over. Thus, reliable automated repair techniques could be beneficial in a production development chain as debugging aids. This finding comforts the human study recently conducted by Tao et al.~\cite{tao_automatically_2014} which suggested that automated repair tools can significantly help debugging tasks.

%\subsubsection*
{\bf On the delay in patch acceptance.} We have observed a delay in the acceptance of tool-supported patches by maintainers. However, given the differences in change operations with fully manual patches, it is likely the case that tool-supported patches are fixing
less severe bugs, which makes their integration a less crucial issue for maintainers. 

Furthermore, negative percentages in evolution gap between submission and acceptance (cf. Figure~\ref{fig:lkml}) suggests that there are many HMG patches that are integrated into the mainline code base without being discussed by maintainers. This finding implies that once the fix pattern has been validated, patches appear to be accepted systematically.

%\subsubsection*
{\bf On the nature of bugs being fixed.} The study of patch locality shows results that are in line with a previous study~\cite{zhong_empirical_2015} which revealed that most fix patches only change a single file. Nevertheless, we have found that, in practice, even tool-supported patches, in a large majority, modify several lines to fix warnings by bug detection tools (which, by the way, generally flag a single line in the code). Although patch size does not, by any means, imply ease of realization, our results suggest that there are considerable numbers of repair targets and shapes that automated repair should aim for.

It is also noteworthy that the spread of change operations over several files may 
carry different implications for the patch construction processes. For example, while a coccinelle
patch may be applying the same change pattern over several files to fix an API function usage, a human patch modifying several files may actually carry data and behavior dependencies among the changes.

%\vspace{-0.3cm}
\subsection{Exploiting Patch Redundancies}
A large body of the literature on program repair has discussed findings
on the repetitiveness/redundancy of code changes in real-world software development~\cite{barr_plastic_2014,nguyen_recurring_2010}. Unfortunately, such findings are
not readily actionable in the context of automated repair since they do not come with insights on how such redundant patches will be leveraged in practice. Indeed,
although it is possible to abstract redundant patches to recommend bug fix
actions~\cite{BissyandeArxiv}, only a few research directions manage to
contextualize them, to some extent, for repair scenarios~\cite{long_analysis_2016}. 
Actually, researchers discuss such redundancies for enriching the repair
space with change operations that are more likely to be appropriate fix operations.

With this study, we see concrete opportunities for exploiting patch redundancies
for systematically building patches and applying (or recommending) them to a specific
identified and localized buggy piece of code. Indeed, bug detection tools, which are
used by various developers who then craft fixes based on specific warnings, and patch application tools, which are based on fix patterns, can be leveraged in an automated repair chain.
The former will be used in the bug detection and localization steps while the latter
will focus on building concrete patches based on patterns found in a database of human fixes created to address warnings by bug detection tools.

To demonstrate the feasibility of this research direction, we have conducted a study
for searching redundancies in patches constructed following warnings by bug detection tools, and investigating the possibility of producing a generic patch which could have been used to derive these concrete patches. Nevertheless, although generic patch inference has been a very fertile research direction in the
past~\cite{andersen_generic_2008,Andersen:2012:SPI:2351676.2351753,meng_sydit:_2011,meng_lase:_2013},
we have experimented available tool supports and found that they do not scale in practice.
We have thus devised a process to split the set of patches into clusters, each containing
patches presenting similar change operations. Figure~\ref{fig:mutation-operations} depicts
the overall process. Based on GumTree sequences of change operations, we rely on a sequential pattern mining tool to extract maximal sequential patterns. We use
a fast implementation of VMSP~\cite{fournier-viger_vmsp:_2014} to find recurrent change patterns at the level of the abstract change operations expressed in Section~\ref{sec:rq4}.
Then, we build clusters of patches based on the elicited patterns, and leverage
SpDiff~\cite{andersen_generic_2008} to attempt the inference of a unique SmPL patch which could instantiate the common redundant concrete repair actions performed in the patches.

With this process, starting with a set of 571 DLH patches, we were able to build
37 clusters based on change operations patterns. Among the clusters, 10 led to
the generation of a common generic patch. We then manually investigated the commit messages associated with the patches in clusters that produced a generic patch, and found that
they indeed largely dealt with the same bug type. This final check confirms, to some extent, the potential to
collect fix patterns from human repair processes to build an automated repair chain leveraging bug detection tools.

%\tb{@Anil: give stats on the change operation patterns, and the inferred patches. Then we provide an example of inferred SmPL patch}

\subsection{Threats to Validity}
We have identified the following threats to validity to our study:
{\em External validity --} We focus on Linux only. It is, however, one of the largest development project, one of the most diverse in terms of developer population, with a significant history for observing trends, and implementing strict patch submission guidelines that try to systematize the tracking of change information. To the best of our knowledge, Linux is the best candidate for observing various patch construction processes, as it encourages the use of tools for bug detection and patching.\\
{\em Construct validity --} We rely on a number of heuristics to collect and process our datasets. We have nevertheless, by design, chosen to be conservative in the way we collect patches in each process with the objective of having reliable and distinctive sets for each process, to further enable replication.\\
{\em Internal validity --} The metrics that we leverage to elicit the differences among the different processes may lead to biased results. However, those metrics were also used in the literature. %\tb{Say less negatively...} Our use of AST differencing tools, and the abstraction of change operations, may also introduce some threats since  GumTree has some limitations: it often fails to parse some nodes and, in rare cases, it indicates a change operation for nodes when there is none

\section{Related Work}
\label{sec.related}

\subsection{Program Repair}

\subsubsection{Studies on Human-Generated Patches}
Studies on patches, generated by human developers, focus on investigating existing patches fully written by developers (i.e., \manual patches) rather than devising a new technique.
Pan et al. explored syntactic bug fix patterns in seven Java projects~\cite{pan_toward_2008}. This study extracted 27 bug fix patterns. 
%The authors presented an automatic pattern extraction tool, leveraging syntactic structures of bug fixes. They figured out that there are similar bug fix patterns across different projects. 
%On the other hand, other researchers investigated human-generated patches to assess automated repair techniques. 
Martinez and Monperrus identified common program repair actions (patterns)~\cite{martinez_mining_2015}, and Zhong and Su reported statistics on 9,000 real bug fixing patches collected from Java open source projects~\cite{zhong_empirical_2015}.
These studies examined features of real bug fixes against whether automated repair techniques can be applied to fix those bugs. 
%The study results pointed out the limitations of automated techniques with respect to fault localization, search space, and multi-line faults. 
%focused on the size of potential search space when applying automated repair techniques to real programs. Their findings show that the likely-correct repair actions are localized and concentrated in a specific subset of the search space (i.e., a smart search strategy can significantly improve the effectiveness of automated repair techniques). 
In addition, Barr et al. formulated a hypothesis called ``\emph{plastic surgery hypothesis}''~\cite{barr_plastic_2014}. They studied how many changes can be graftable by using snippets that can be found in the same code base where the changes are made. 
%The hypothesis must be examined since many repair techniques rely on genetic programming. 

%This implies that snippets are able to be found in the search space (i.e., the same source code files or modules where a bug exists) even if they are not exactly in the same shape with the desired patches. The study revealed that 43\% of changes are graftable from the same codebase of the changes.

%\dongsun{+ discussion and comparison with our study results.}
%\vspace{-3mm}
\subsubsection{Studies on Tool-aided Patches}

As discussed in Sections~\ref{sec.introduction} and~\ref{sec.dataset}, generating tool-aided patches indicates that developers create program patches with an aid of tools, rather than generating patches from scratch. 
%This line of studies examines how patch-aid tools can help developer generate correct patches.
Tao et al. supposed that automated repair tools can provide aids to debugging tasks~\cite{tao_automatically_2014}. They adopted \patternfix~\cite{kim_automatic_2013} as a patch recommendation tool and gave patches generated by the tool to experiment participants. The findings include that automatically generated patches can significantly help debugging tasks.
MintHint~\cite{kaleeswaran_minthint:_2014} is a semi-automatic repair technique, which can help developer find correct patches. This technique does statistical correlation analysis to locate program expressions likely to perform repaired program executions. 
%It then generates repair hints based on pattern-matching. The authors conducted a user study to evaluate MintHint and the study results show that the technique substantially improved developer productivity.

%Wei et al. proposed a contract-based patch generation
%technique~\cite{wei_automated_2010}. This technique also relies on specifications
%(i.e., contracts) so that it cannot generate patches fully automatically. Based on the specifications, the technique can locate
%potential defects. This technique can generate four kinds of
%program variants and check out whether they violates the specification. If a variant eliminates a corresponding violation, it can be a patch. 

%\dongsun{+ discussion and comparison with our study results.}
%\vspace{-3mm}
\subsubsection{Automated Patch Generation}

Generating patches with automated tools implies minimizing a developer's effort in debugging. It often indicates that fully automated procedures including fault localization, code modification, and patch verification. Recent endeavors achieved an impressive progress as follows.

%To automatically fix program bugs, several techniques have been proposed. Although these techniques leverage many different ideas and algorithms such as evolutionary computing and constraints solving, their common goal is to minimize human effort for program repair and reduce the time and cost for debugging. One group of techniques focuses on a specific type of bugs such as software crashes. On the other hand, other techniques attempt to fix a wide range of bugs.

%Arcuri et al. introduced an automatic patch generation
%technique~\cite{arcuri_novel_2008}.
%This approach leverages genetic programming~\cite{koza_genetic_1992} to mutate programs with respect to the ration of passing test cases (i.e., fitness function). The approach can also use
%formal specifications to compute a fitness function. Although this approach was the first attempt to apply an evolutionary algorithm to automated program repair, its scalability and applicability were not confirmed since
%their evaluation was limited to
%small programs such as bubble sorting and triangle classification.

Weimer et al.~\cite{weimer_automatically_2009} proposed GenProg, an automatic patch generation technique based on genetic programming~\cite{koza_genetic_1992}. This technique randomly mutates buggy statements to generate several different program variants that are potential patch candidates. 
%The variants are verified by running both passing and failing test cases. If a variant passes all test cases, it is regarded as a successful patch of a given bug. 
In 2012, the authors extended their previous work by adding a new mutation operation, replacement and removing the switch operation~\cite{le_goues_genprog:_2012}. 
%This improved technique successfully fixed 55 out of 105 real bugs~\cite{le_goues_genprog:_2012}. 
SemFix~\cite{nguyen_semfix:_2013} leverages program synthesis to generate patches. The technique assumes that buggy predicates are an unknown function to be synthesized. 
%Then, it formulates the unknown function as a set of constraints to solve. Those constraints are given to a constraint solver. Based on the result of the solver, this technique generates a repaired predicate passing all the given test cases. 
The technique is successful for several bugs, but it is only applicable to ``one-line bug'', in which only one predicate is buggy. DirectFix~\cite{mechtaev_directfix:_2015} and Angelix~\cite{mechtaev_angelix:_2016} extended Semfix so that it can generate patches for bugs in larger and complex (w.r.t the search space) programs in a simpler way.
%Moreover, it is effective only for buggy predicates that a constraint solver can address. On the other hand, the repair technique of this project is not limited to ``one-line'' bugs.
PAR~\cite{kim_automatic_2013} automatically generates patches by using fix patterns learned from human-written patches. This technique is inspired by the fact that patches are redundant. 
%In other words, there are common fix patterns in human-written patches. For example, null-checkers are commonly added for bugs throwing a null pointer exception. By leveraging fix patterns, the authors of this technique created 10 fix patterns, which are program edit scripts. The fix patterns mimic human developers who create commonly applied patches. 
%Compared to other techniques, this technique can generate more acceptable patches while mutation-based techniques such as GenProg may make nonsensical patches that developers can reject due to maintenance problems.

%\dongsun{+ discussion and comparison with our study results.}
%\vspace{-12pt}

\subsection{Patch Acceptability}

%Although many automated repair techniques successfully generate correct patches (often verified by test cases), the techniques are often faced with an intrinsic problem; developers sometimes refuse to accept and merge the generated patches due to lack of readability or maintainability in spite of their functional correctness. In particular, techniques relying on genetic mutation often suffer from nonsensical patch issues. 

Fry et al. conducted a human study to indirectly measure the quality of patches
generated by \genprog by measuring patch maintainability~\cite{fry_human_2012}.
They presented patches to participants and asked maintainability related
questions developed by Sillito et al.~\cite{sillito_questions_2006}.
%In addition, they presented machine-generated change
%documents~\cite{buse_automatically_2010} along with patches to participants.
They found that machine-generated patches~\cite{le_goues_genprog:_2012} with
machine-generated documents~\cite{buse_automatically_2010} are comparable to
human-written patches in terms of maintainability.
\patternfix~\cite{kim_automatic_2013} is presented to deal with nonsensical patches. The approach generates patches based on fix patterns, which are learned from human-written patches. The fix patterns generalize common repair actions from more than 60,000 real bug fixes enabling \patternfix to avoid generating nonsensical patches.

%\dongsun{+ discussion and comparison with our study results.}
%\vspace{-12pt}

\subsection{Program Matching and Transformation}

%Automated program matching and transformation can help program repair and patching. Even if a repair technique could reason about a correct behavior resolving a defect, applying the behavior to actual source code is completely different story. Since source code consists of a complex hierarchical structure, precise context matching and transformation techniques are essential for a successful repair.

SYDIT~\cite{meng_sydit:_2011} automatically extracts an edit script from a program change. 
In its scenario, a user must specify the program change to extract the edit script from.
Coccinelle~\cite{Brunel:2009:FFP:1480881.1480897}, on the other hand, directly lets the user
specify the edit script in a user-friendly language, and performs the transformation by matching the change pattern with code context. It has been used in several debugging tasks in the literature~\cite{bissyande2012diagnosys,bissyande2014ahead,bissyande2013contributions,BissyandeArxiv,Palix:2011:FLT:1950365.1950401}. LASE~\cite{meng_lase:_2013} differs from SYDIT as it can generate a generalized edit script based on multiple changes of Java programs.
Another approach in this direction is SpDiff~\cite{Andersen:2012:SPI:2351676.2351753,andersen_generic_2008} supports the extraction of a subset of common changes (i.e., SmPL patches that are fed to Coccinelle) from several concrete patches.
%However, the users of LASE still must specify existing changes to fix a specific bug, which is not realistic in practice.
%
%\tb{Coccinelle must be mentionned here}
%
%\tb{We should also talk about generic patch inference -- Jesper Anderson}

%\input{discussions}
%\vspace{-3mm}
\section{Conclusion}
\label{sec.conclusion}

We have studied the impact of tool support in patch construction, leveraging 
real-world patching processes in the Linux kernel development project. 
We investigated the acceptance of tool-supported patches in the development 
chain as well as the differences that may exist in the kinds of bugs that such 
patches fix in comparison with traditional all-hand written patches. We show 
that in the Linux ecosystem, bug detection and patch application tools are 
already heavily used to unburden developers, and already enable relatively 
complex repair schema, contrasting with a number of repair approaches 
in the state-of-the-art literature of automated repair. 
An artefact dataset on this study is available at \url{https://goo.gl/f1mRMM}.
%\vspace{-6mm}
\section*{Acknowledgements}
The authors would like to thank the anonymous reviewers for their helpful comments and suggestions. This work was supported by the Fonds National de la Recherche (FNR), Luxembourg, under projects RECOMMEND C15/IS/10449467 and FIXPATTERN C15/IS/9964569.

%\begin{CCSXML}
%<ccs2012>
%  <concept_id>10010520.10010553.10010562</concept_id>
%  <concept_desc>Computer systems organization~Embedded systems</concept_desc>
%  <concept_significance>500</concept_significance>
%  <concept_id>10010520.10010575.10010755</concept_id>
%  <concept_desc>Computer systems organization~Redundancy</concept_desc>
%  <concept_significance>300</concept_significance>
%  <concept_id>10010520.10010553.10010554</concept_id>
%  <concept_desc>Computer systems organization~Robotics</concept_desc>
%  <concept_significance>100</concept_significance>
%  <concept_id>10003033.10003083.10003095</concept_id>
%  <concept_desc>Networks~Network reliability</concept_desc>
%  <concept_significance>100</concept_significance>
%</ccs2012>  
%\end{CCSXML}

%\ccsdesc[500]{Computer systems organization~Embedded systems}

\balance
\bibliographystyle{ACM-Reference-Format}
\bibliography{bib/paper} 

\end{document}